%% file: Aurora-MFPrev2.tex
\documentclass[usenatbib]{mn2e}
\usepackage{amssymb,amsmath,graphicx,natbib,setspace,pifont}
\newcommand{\HI}{\hbox{{\sc H}{\sc i}} }
\newcommand{\HII}{\hbox{{\sc H}{\sc ii}} }

\newcommand{\NHI}{{N_{\rm HI}}}

\newcommand{\cmsq}{\,{\rm cm^{-2}}}
\newcommand{\cms}{\,{\rm cm^{2}}}
\newcommand{\cmcb}{\,{\rm cm^{-3}}}

\newcommand{\nH}{n_{\rm _{H}} }

\newcommand{\nHI}{n_{\rm _{HI}}}

\newcommand{\Ob}{\Omega_{\rm b} }
\newcommand{\Om}{\Omega_{\rm m} }

\newcommand{\Ol}{\Omega_{\Lambda} }
\newcommand{\ns}{n_{\rm s} }
\newcommand{\sigeight}{\sigma_{\rm 8} }

\newcommand{\Msun}{{\rm M_{\odot}} }

\newcommand{\cMpch}{h^{-1} {\rm cMpc} }

\newcommand{\Gadget}{{\small GADGET-3} }

\newcommand{\TRAPHIC}{{\small TRAPHIC}~}

\newcommand{\apjl}{{ApJ}}
\newcommand{\aj}{{AJ}}
\newcommand{\apj}{{ApJ}}
\newcommand{\apjs}{{ApJS}}
\newcommand{\aap}{{A\&A}}
\newcommand{\mnras}{{MNRAS}}

\begin{document}

\title[The MFP during the EoR]{The mean free path of hydrogen ionizing photons during the epoch of reionization}

\author[A.~Rahmati \& J.~Schaye] {Alireza~Rahmati$^{1}$\thanks{ali.rahmati@gmail.com}\& Joop~Schaye$^{2}$\\
  $^1$Institute for Computational Science, University of Z\"urich, Winterthurerstrasse 190, CH-8057 Z\"urich, Switzerland\\	
  $^2$Leiden Observatory, Leiden University, P.O. Box 9513, 2300 RA, Leiden, The Netherlands\\}

\maketitle

\begin{abstract} 
We use the Aurora radiation-hydrodynamical simulations to study the mean free path (MFP) for hydrogen ionizing photons during the epoch of reionization. We directly measure the MFP by averaging the distance 1 Ry photons travel before reaching an optical depth of unity along random lines-of-sight. During reionization the free paths tend to end in neutral gas with densities near the cosmic mean, while after reionizaton the end points tend to be overdense but highly ionized. Despite the increasing importance of discrete, over-dense systems, the cumulative contribution of systems with $\NHI \lesssim 10^{16.5}~\cmsq$ suffices to drive the MFP at $z \approx 6$, while at earlier times higher column densities are more important. After reionization the typical size of $\HI$ systems is close to the local Jeans length, but during reionization it is much larger. The mean free path for photons originating close to galaxies, $\rm{MFP_{gal}}$, is much smaller than the cosmic MFP. After reionization this enhancement can remain significant up to starting distances of $\sim 1$~comoving Mpc. During reionization, however, $\rm{MFP_{gal}}$ for distances $\sim 10^2 - 10^3$ comoving kpc typically exceeds the cosmic MFP. These findings have important consequences for models that interpret the intergalactic MFP as the distance escaped ionizing photons can travel from galaxies before being absorbed and may cause them to under-estimate the required escape fraction from galaxies, and/or the required emissivity of ionizing photons after reionization.
\end{abstract}

\begin{keywords}
  radiative transfer -- methods: numerical -- galaxies: formation -- galaxies: high-redshift -- intergalactic medium -- dark ages, reionization, first stars.
\end{keywords}

\section{introduction}
\label{sec:intro}

The reionization of hydrogen on cosmological scales is the last and arguably one of the most important major phase transitions in the physical state of the baryonic Universe. Several observables, including the Thompson optical depth for photons from the Cosmic Microwave Background (CMB; e.g., \citealp{Planck16}) and the rapid change in the opacity of the intergalactic medium (IGM) at $z \approx 6$ \citep[e.g.,][]{Fan06}, indicate that hydrogen reionization was complete by the time the Universe was $\approx 1$ billion years old. There are, nonetheless, major unanswered questions about the physical processes governing the epoch of reionization (EoR) and the details of the IGM phase transition from highly neutral to highly ionized.  

A useful quantity for tracing the IGM phase transition is the distance hydrogen ionizing photons typically travel before being absorbed. Well before reionization, this distance is nearly zero and a hypothetical 1 Ry photon is absorbed near its source. After reionization, on the other hand, such a photon can travel relatively large cosmic distances before being absorbed. The distribution of distances 1 Ry photons can travel at different stages of reionization is tightly correlated with the physical state of the IGM and therefore can be used to quantify it. In fact, the mean of the distribution, the mean free path (MFP), is a standard measure for quantifying the opacity of the IGM which is widely used in both observational and in theoretical studies.

The contribution of different $\HI$ systems to the absorption of 1 Ry photons on cosmic scales, and thereby their impact on the MFP of Lyman Limit photons, depends on their $\HI$ column densities as well as on their relative abundances. For the post-reionization Universe, there are robust observational constraints on the column density distribution function of $\HI$ absorbers for a wide range of redshifts \citep[e.g.,][]{Storrie94, Kim02, Songaila10, Prochaska10, Omeara13, Rudie13} which can be used to calculate the contribution of different column densities to the cosmic absorption of hydrogen ionizing photons. The shape of the observed $\HI$ distribution suggests that the average rate of absorption is dominated by optically thick systems, the so called Lyman Limit systems (LLS) with $\NHI > 1.6 \times 10^{17}\cmsq$. The reason for this is that the higher frequency of systems with lower $\HI$ column densities is not high enough to compensate for their smaller opacities \citep[e.g.,][]{Miralda03}. Moreover, the steep decrease in the frequency of $\HI$ systems with higher column densities makes their contributions less important.

Indeed, earlier attempts that used the measured frequency of LLSs to estimate the MFP \citep[e.g.,][]{Meiksin93,Songaila10} are in good agreement with more recent direct measurements of the MFP using the drop in the continuum flux of high-resolution quasar spectra \citep{Prochaska13, Worseck14} up to redshifts $z \approx 5$. At higher redshifts, however, it becomes increasingly more difficult to put stringent constraints on the $\HI$ column density distribution function, partly due to the small number of observed bright quasars at those redshifts which also makes the direct measurements of the MFP hardly possible. Such limitations currently prevent us from measuring the evolution of the MFP and assessing the relative importance of different $\HI$ systems for the cosmic absorption of 1 Ry photons close to the epoch reionization. More importantly, the increasingly high mean Gunn-Peterson opacity of the IGM at redshifts higher than $z\sim 5$ makes studying the spectra of sources at those redshifts and identifying individual absorbers very difficult.

The notion of the MFP as a universal quantity which only depends on time is often used in analytic models to connect the average emissivity of ionizing photons to the large-scale radiation field after reionization. The MFP, however, could be very different close to galaxies compared to its cosmic averaged value. On the one hand, the average ionizing radiation field is expected to be stronger closer to galaxies which can reduce the abundance of $\HI$ absorbers and therefore increases the MFP close to galaxies \citep[e.g.,][]{Rahmati13b}. On the other hand, galaxies reside in environments with densities much higher than a randomly chosen location in the Universe. This may result in an enhanced abundance of $\HI$ absorbers and therefore shortened MFPs close to galaxies. Indeed, observations show that $z\approx2$ star-forming galaxies and quasars reside in regions with enhanced $\HI$ densities \citep[e.g.,][]{Rakic12,Rudie12,Rudie13,Turner14,Prochaska13}, consistent with recent hydrodynamical simulations \citep[e.g.,][]{Rahmati15}.

While observational measurements of the MFP shortly after the EoR await significant progress in our observational capabilities, radiation-hydrodynamical simulations can be used to gain insight into the origin of the MFP. Such simulations should, however, fulfil some specific requirements: they should have sufficient resolution to model both the sources and sinks of ionizing photons \citep[e.g.,][]{McQuinn07,Alvarez12,Emberson13}. Numerical simulations of the EoR have begun to reach a level of maturity that allows them to fulfil those requirements while reproducing simultaneously the timing of reionization and the observed properties of the galaxy population during the EoR \citep{Gnedin14, Pawlik15, Pawlik16}. 

In this paper we use the Aurora simulation suite \citep{Pawlik16} to investigate the evolution of the MFP and its dependence on the distance from galaxies during the EoR together with the nature of the $\HI$ absorbers that dominate the opacity of the IGM at that epoch. Aurora consists of radiation hydrodynamical simulations of the co-evolution of galaxies and the IGM during the first $\approx 1$ billion years of the Universe. The simulations have box sizes of up to $50h^{-1}$ comoving Mpc (cMpc) combined with sub-kpc spatial resolutions. The parameters that regulate the key sub-grid physical processes, i.e., the supernova feedback and sub-resolution escape of ionizing photons in the interstellar medium (ISM), are calibrated such that the Aurora simulations with different box sizes and resolutions have similar reionization histories and galaxy populations, in excellent agreement with available observational constraints. Both features are crucial for studying the evolution of the IGM and MFP during the EoR while the available variations in the captured volume and resolutions allow us to robustly estimate numerical uncertainties. In particular, as we explicitly show in this paper, the sub-kpc resolution of the simulation guarantees that the main $\HI$ sinks of ionizing radiation, namely LLSs with expected sizes $1-10$ proper kpc \citep{Schaye01,Rahmati13a, Erkal15} are well resolved. 
  
The structure of this paper is as follows. In $\S$\ref{sec:method} we discuss the design of the Aurora simulations and our methodology for analysing them. Section $\S$\ref{sec:results} contains the main results which cover the evolution of the MFP during the EoR ($\S$\ref{sec:MFP}), the evolution of the distributions of free paths ($\S$\ref{sec:FPs}) the evolution of different physical properties of the $\HI$ absorbers ($\S$\ref{sec:ODUs}), their size distributions ($\S$\ref{sec:HIsizes}), their contribution to the MFP ($\S$\ref{sec:MFPorigin}) and the variations in the MFP close to galaxies ($\S$\ref{sec:GalMFP}). We summarize the paper in $\S$\ref{sec:dend}. 

\section{Methodology}
\label{sec:method}
\subsection{Simulations}
\input{simparam}
We use the Aurora suite of radiation-hydrodynamical simulations of galaxy formation during the epoch of reionization. We briefly explain the main components of the simulation design in the following, but invite the interested reader to find a detailed description in \citet{Pawlik16}. 

A modified version of the \Gadget code (last described in \citealp{Springel05}) coupled with the radiative transfer code \TRAPHIC \citep{Pawlik08,Pawlik11} was used to perform the simulations. Note that \TRAPHIC performs the readiative transfer at the native spatially adaptive resolution of the hydrodynamics code. The cosmological parameters used for the simulations are: $\{\Om=0.265,\ \Ob=0.0448,\ \Ol=0.735,\ \sigeight=0.801,\ \ns=0.963,\ h=0.71\}$ \citep{Komatsu11}. The box sizes range from $12.5$ to $50~\cMpch$ and gas (dark matter) particle masses are in the range $2.55\times10^5$ - $1.63\times10^7~\Msun$ ($1.25\times10^6$ - $8.2\times10^7~\Msun$), as listed in Table \ref{tbl:sims}. The pressure-dependent model of \citet{Schaye08} is used as the sub-grid prescription for star formation. Stellar evolution assumes a \citet{Chabrier03} initial mass function and determines the element-by-element gas enrichment \citep{Wiersma09}, the rates of Supernova (SN) events, and stellar ionizing luminosities \citep{Schaerer03}. Stellar (SN) feedback is implemented thermally following the stochastic method of \citet{Dalla-Vecchia12} with $\Delta \rm{T = 10^{7.5}~K}$. Because the simulations lack the resolution to predict the radiative losses in the ISM from first principles, different fractions of the total available SN energy (ranging from 0.6 to 1.0) are used for simulations with different resolutions such that they all result in very similar star-formation-rate functions at redshift $z = 7$. 

A subresolution escape fraction, $f_{\rm{esc}}^{\rm{subres}}$, is adopted to account for absorption of photons in the unresolved part of the interstellar medium, and resolution dependent differences in star formation. The parameter $f_{\rm{esc}}^{\rm{subres}}$ is calibrated for simulations with different resolution to force them to reach a volume-weighted neutral hydrogen fraction of $0.5$ at $z \approx 8.3$. The required $f_{\rm{esc}}^{\rm{subres}}$ values for the low, reference and high resolutions are 1.3,  0.8 and 0.6, respectively. However, we also use a simulation with a resolution identical to the reference simulation, but with a lower value $f_{\rm{esc}}^{\rm{subres}} = 0.3$, \emph{L12N256-L}, such that it reaches the same neutral fraction only at $z \approx 7$. As we showed in \citet{Pawlik16}, the latter run produces a Thompson optical depth of $\tau_{\rm{reion}} = 0.055$ towards the CMB, which is in better agreement with the latest Planck measurements \citep{Planck16} compared to our reference model with $\tau_{\rm{reion}} \approx 0.068$, which was very close to the previous Planck measurements \citep{Planck15}. It is, however, important to note that all simulations produce Thompson optical depths that are consistent with both of the aforementioned Planck measurements within their reported uncertainties.

As we showed in \citet{Pawlik16}, the simulations also match various other observational constraints, including the evolution of the star formation rate density, the evolution of star formation rate function and the evolution of the neutral fraction and background photoionization rate after reionization.

\subsection{Mean free path calculation}
\label{sec:MFPcalc}
\begin{figure*}
\centerline{\hbox{\includegraphics[width=0.5\textwidth]
             {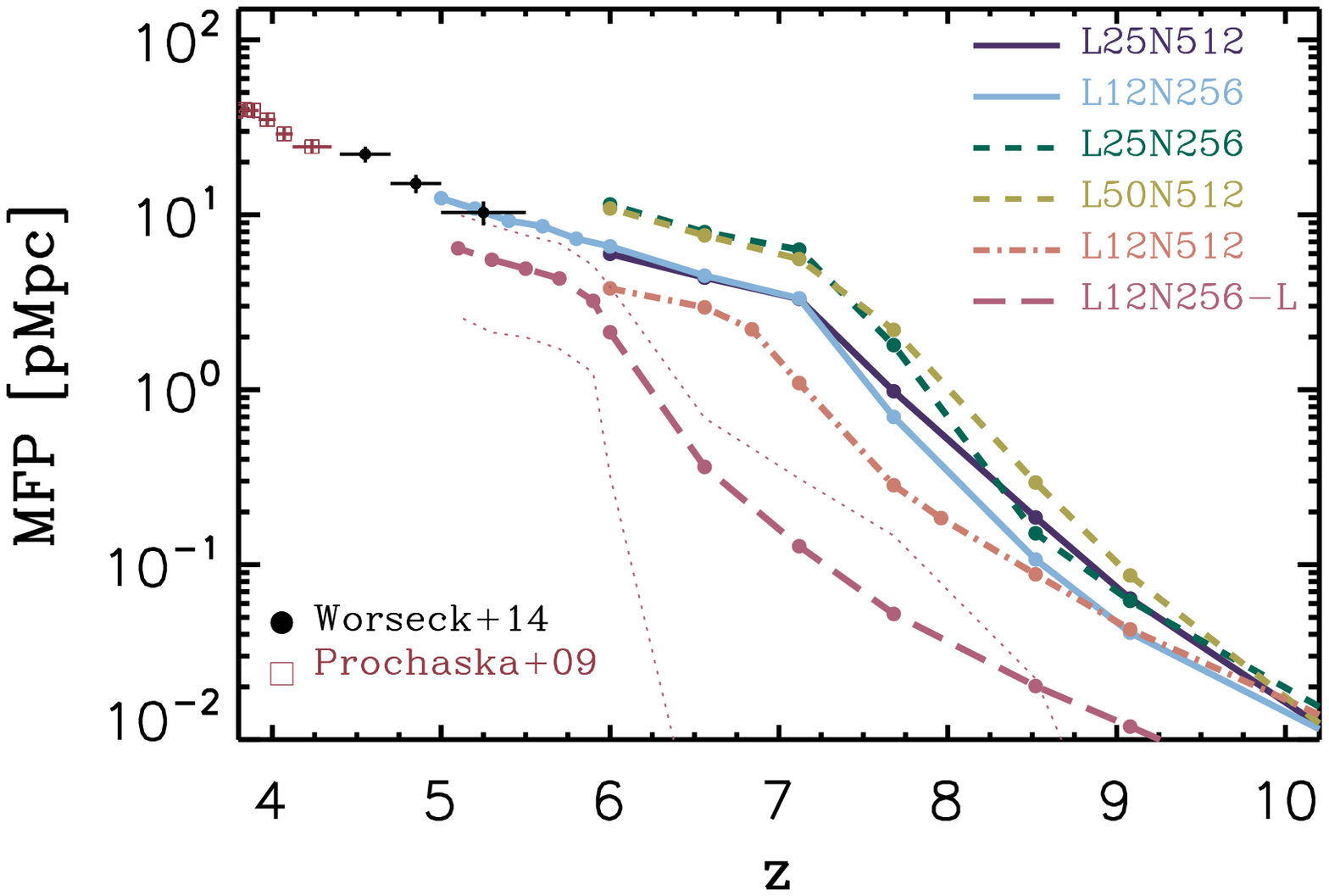}}
             \hbox{\includegraphics[width=0.5\textwidth]
             {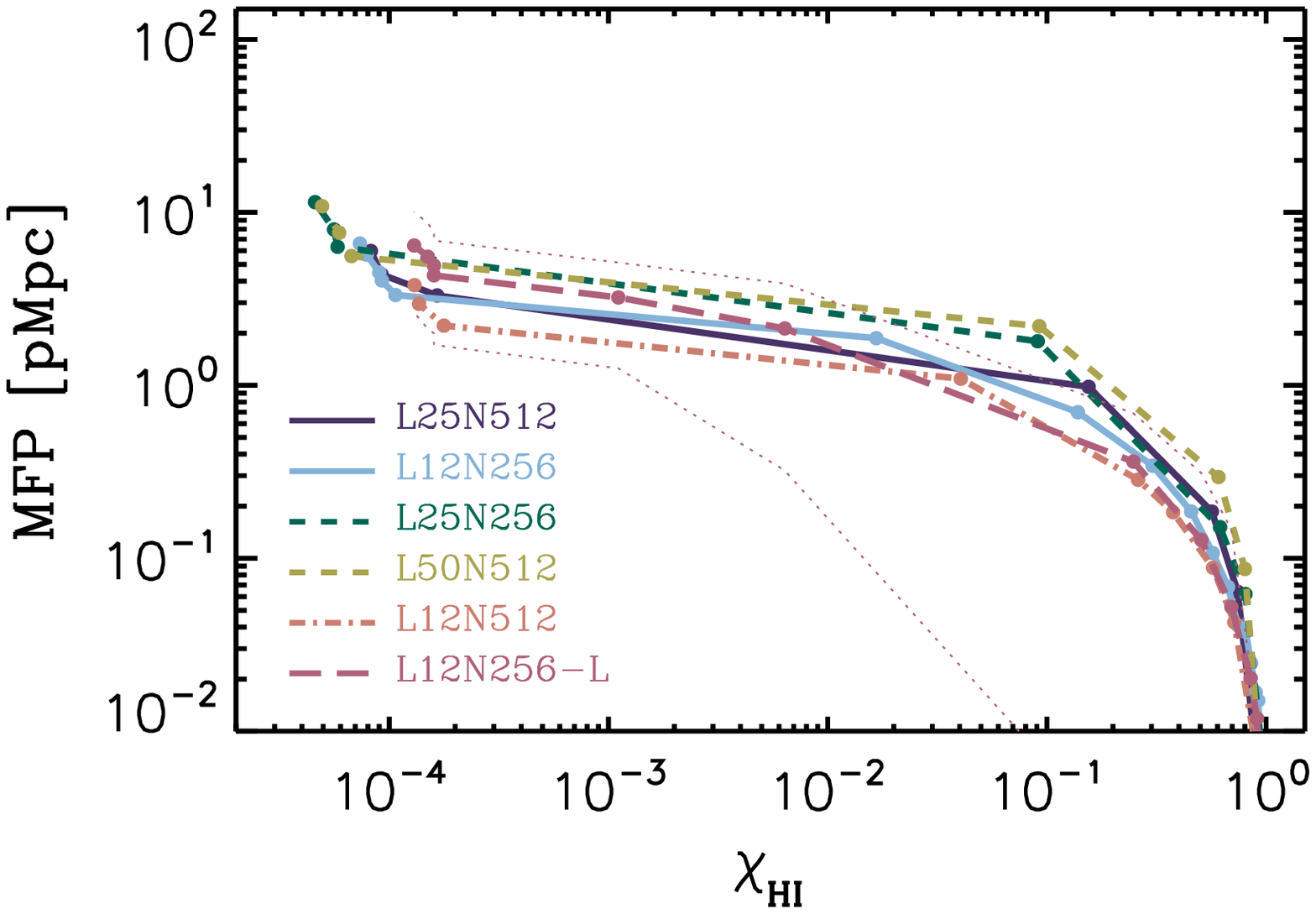}}}
\caption{The MFP evolution in different Aurora simulations as a function of redshift (left; using proper units) and as a function of the volume-weighted $\HI$ fraction (right). To indicate the scatter around the MFP, the 15th-85th percentiles of the free-path distribution in the \emph{L12N256-L} simulation are shown using dotted curves. The observational measurements of the MFP from \citet{Worseck14} and \citet{Prochaska09} are shown in the left panel (symbols with error bars) and agree well with our model predictions, particularly those from the reference model used in the \emph{L12N256} and \emph{L25N512} simulations. The MFP evolution in the simulations with similar reionization histories is very similar. The differences in the MFPs in the different simulations is mainly due to their different IGM $\HI$ fractions and hence become small at fixed $\HI$ fractions.}
\label{fig:MFP-evol}
\end{figure*}
We directly calculate the MFP for LL photons, i.e., photons with $1~\rm{Ry}$ energy, equivalent to a wavelength of $912 \AA$, by generating a large number of lines-of-sights (LOS) through the simulation volume and measuring, along each LOS, the distance required to reach an optical depth of unity:
\begin{equation}
\tau_{912} = \NHI \times \sigma_{912} = 1,
\label{eq:tau}
\end{equation}
 where $\sigma_{912}  = 6.28 \times 10^{-18}~\cms$\citep{Osterbrock06}. We call this distance the ``free-path" (FP) along the LOS. Noting that our LOS have lengths identical to the length of the simulation box, if an optical depth of unity is not reached by the end of a LOS, we randomly draw another LOS and continue the integration until an optical depth of unity is reached. Then, by averaging over all FPs for a given simulation at a given time, we calculate the MFP. We typically use 3600 LOS and a $\approx 3-10$ ckpc spatial resolution along the LOS for each MFP measurement. We verified that these numerical choices result in converged results.

The definition described above for the FP and MFP assumes mono-frequency radiation and therefore is different from the average distance a typical hydrogen ionizing photon travels before reaching an optical depth of unity. In practice, the energy of a typical hydrogen ionizing photon depends on the nature of the sources of ionizing radiation and is higher than $1~\rm{Ry}$. This means that the MFP of typical hydrogen ionizing photons, which see smaller effective absorption cross-sections, is longer than what is calculated based on equation \eqref{eq:tau}. However, this definition is useful for quantifying the distribution of neutral hydrogen in the IGM and is widely used in both theoretical \citep[e.g.,][]{Miralda03,GnedinFan06,Kuhlen12} and observational studies \citep[e.g.,][]{Prochaska09, Omeara13, Rudie13,Worseck14}.

We also assume that the frequency of LL photons does not change as they travel along the paths over which we calculate their extinction. This is, however, a fair assumption given that here we study the evolution of the MFP during the EoR where LL photons are absorbed on relatively short distances before experiencing significant cosmological redshifting.

\section{Results}
\label{sec:results}
\subsection{Mean free path}
\label{sec:MFP}

The evolution of the proper MFP in the different Aurora simulations is shown in the left panel of Figure \ref{fig:MFP-evol}. The evolution is qualitatively similar for all the simulations: the MFP increases rapidly (with time) during the epoch of reionization when the IGM $\HI$ fraction decreases rapidly. At later stages, when the IGM is ionized to very low $\HI$ fractions ($\sim 10^{-4}$), the MFP evolves more slowly. For the Aurora simulations with identical reionization times (i.e., all but the \emph{L12N256-L} simulation), the MFPs before reionization ($z > 8$) increases slightly with the simulation box size. At later times, on the other hand, the MFP becomes insensitive to the box size but decreases with the resolution. This trend can be understood by noting that the early evolution of the MFP is driven by the typical size of the $\HII$ regions (see also \citealp{GnedinFan06}), which is larger in larger simulations. This is a consequence of the increase in the size of massive collapsed structures as the simulation volume increases. At later times, when the $\HII$ regions overlap, the MFP is more sensitive to the abundance of IGM $\HI$ absorbers and the photoionization rate seen by them\footnote{Note that the comparison between the late stages of the MFP evolution in simulations with identical resolutions but different box sizes confirms the robustness of our procedure for measuring the MFP when it exceeds the box size.}. As shown by \citet{Pawlik16}, the IGM photoionization rates of our simulations with different resolutions differ from one another. This is mainly due to resolution-dependent differences in the fraction of the radiation that escapes from galaxies into the IGM. While we minimize those differences at $z \approx 8.3$ by calibrating the $f_{\rm{esc}}^{\rm{subres}}$ parameter, they evolve (e.g., through an evolution in the ratio between the total star-formation rate densities of the different runs). This translates into a resolution-dependent MFP where lower-resolution simulations yield higher IGM photoionization rates, and therefore a lower average $\HI$ absorption, resulting in longer MFPs at the later stages of reionization. 

The predicted MFPs are in a reasonable agreement with observational constraints from \citet{Worseck14} and \cite{Prochaska09} as shown in the left panel of Fig. \ref{fig:MFP-evol}. Excellent agreement is achieved for our reference model \emph{L25N512} and for \emph{L12N256}. However, there is a rather large scatter in the distribution of FPs, which is shown using dotted curves for \emph{L12N256-L} only. Considering this scatter, which is very similar for all of our simulations, the results from all the simulations shown in Fig. \ref{fig:MFP-evol} may be consistent with the observational measurements. 

The evolution of the MFP as a function of the volume-weighted mean $\HI$ fraction is shown in the right panel of Figure \ref{fig:MFP-evol}. The greater similarity between the MFP evolutions of the \emph{L12N256} and \emph{L12N256-L} simulations indicates that the differences seen in the left panel of Figure \ref{fig:MFP-evol} are mainly due to differences in the ionization state of the IGM and hence in the timing of reionization in the two simulations. The MFP increases rapidly as a function of $\HI$ fraction at the early stages of reionization, when it increases by a factor of $\gtrsim 1000$ between the $\HI$ fractions of $\chi_{\rm{HI}} = 0.9$ and $0.1$. During a very short period of time after the completion of reionization (i.e., $6 < z < 6.5$ for \emph{L12N256-L} and $ 7 < z < 7.5$ for all the other simulations), the MFP remains nearly constant while the $\HI$ fraction decreases by a factor of $\gtrsim 1000$ to its final value of $\lesssim 10^{-4}$. The subsequent increase in the proper MFP with decreasing IGM $\HI$ fraction steepens because the cosmic expansion is causing a continuous decrease in the abundance of $\HI$ absorbers and the average density of the IGM.

\begin{figure*}
\centerline{\hbox{\includegraphics[width=0.5\textwidth]
             {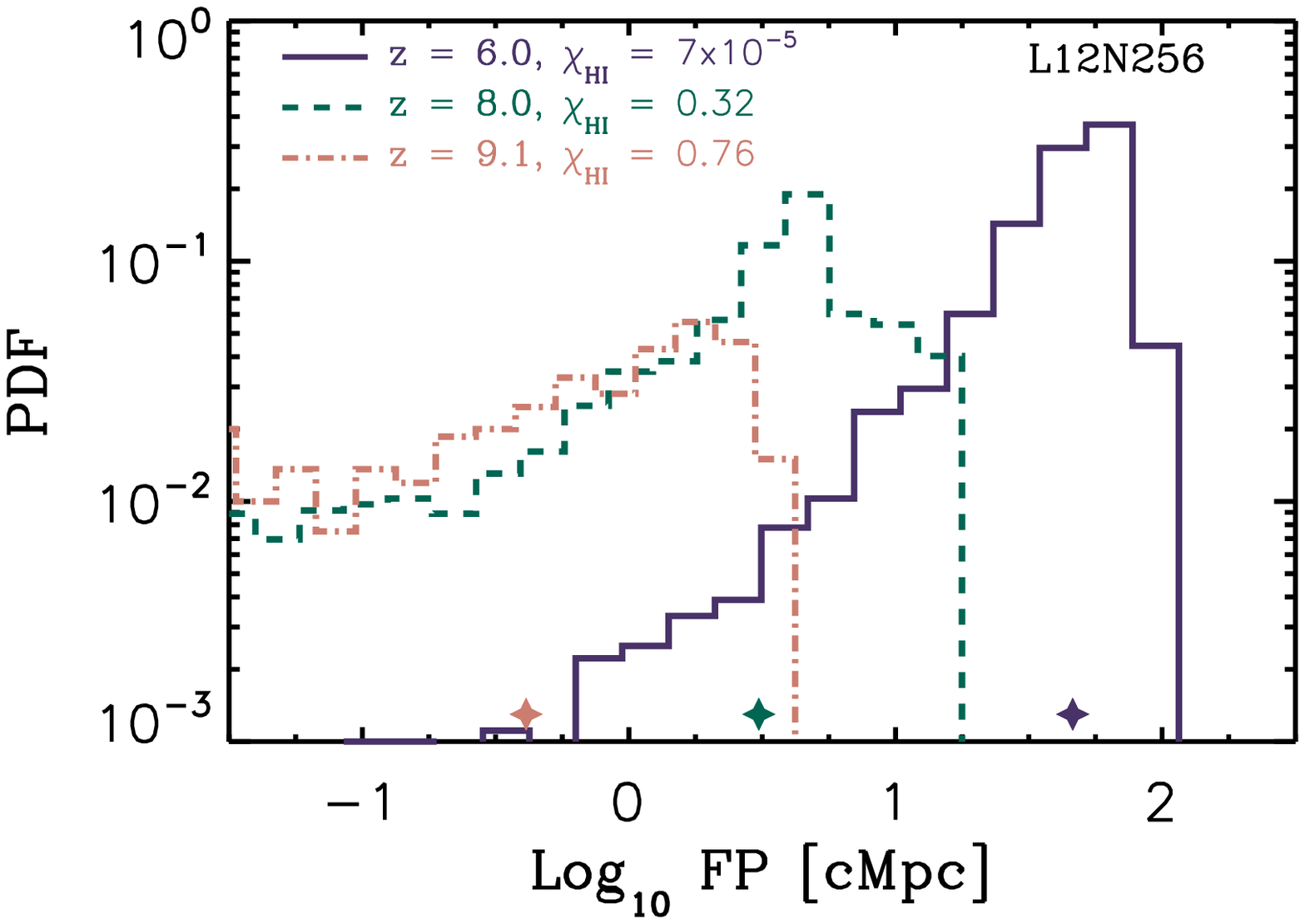}}
             \hbox{\includegraphics[width=0.5\textwidth]
             {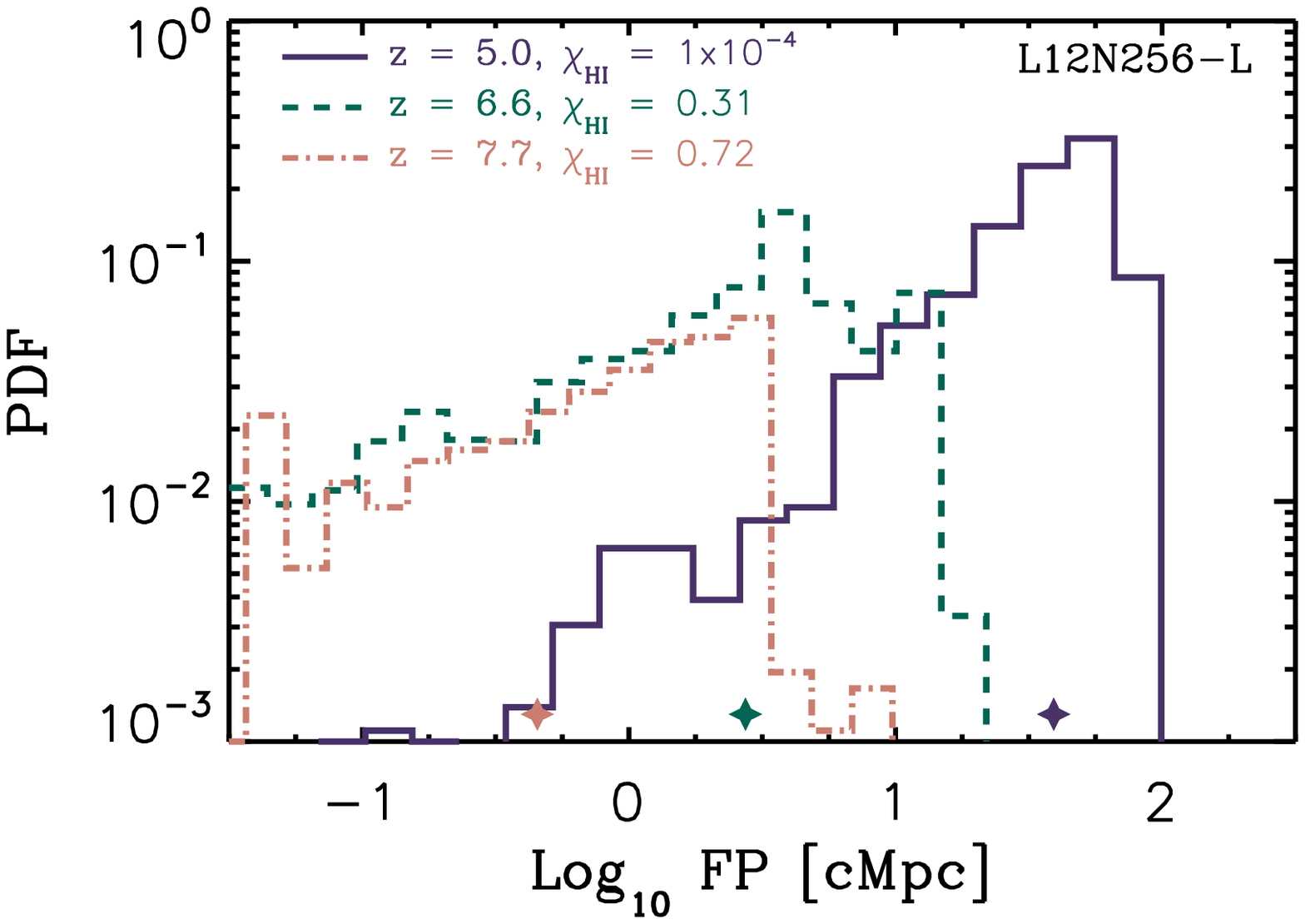}}}
\caption{Probability distribution function of distances required for $912 \AA$ photons to reach an optical depth of unity along random lines-of-sight. Left and right panels show the results for the   \emph{L12N256} and \emph{L12N256-L} simulations. In each panel, the distribution of free-path (FP) lengths is shown at three different times where the MFP of a given distribution function is shown by a star near the bottom of each plot, using the color matching the corresponding distribution. The 3 different redshifts in the two panels are chosen to roughly correspond to the same stages of reionization, resulting in similar MFPs and FP distributions.}
\label{fig:FP-dists}
\end{figure*}
\begin{figure*}
\centerline{\hbox{\includegraphics[width=0.5\textwidth]
             {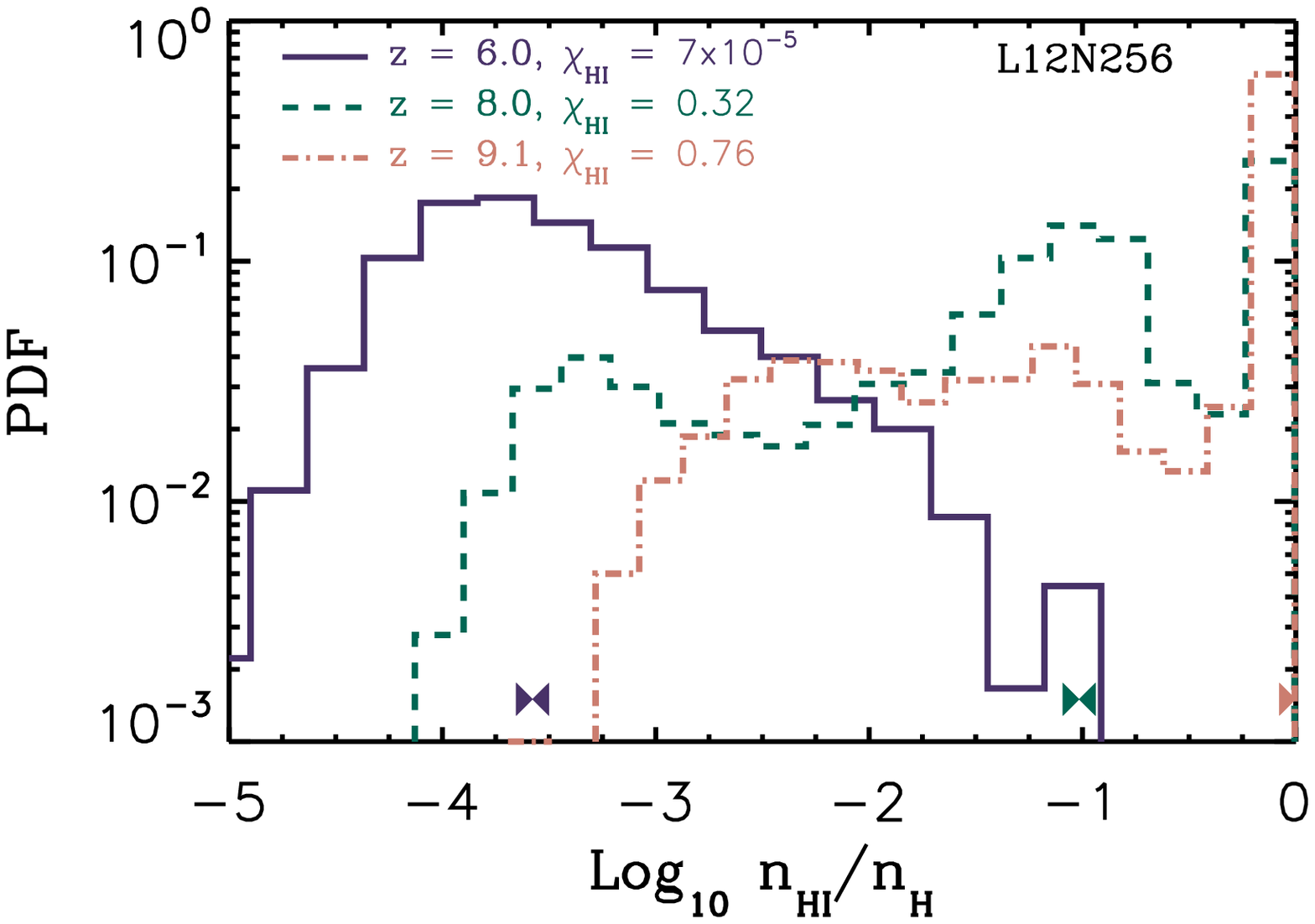}}
             \hbox{\includegraphics[width=0.5\textwidth]
             {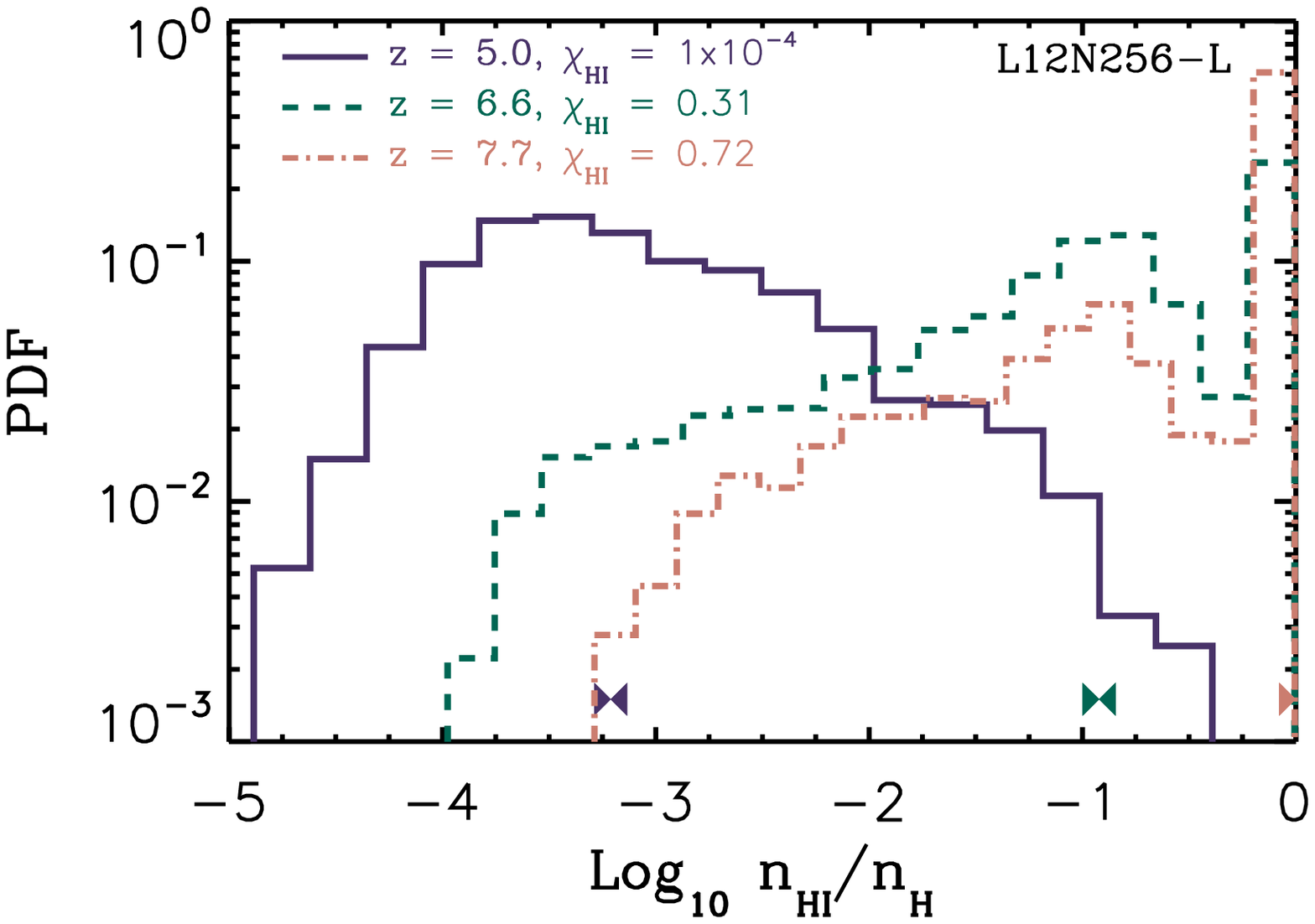}}}
\centerline{\hbox{\includegraphics[width=0.5\textwidth]
             {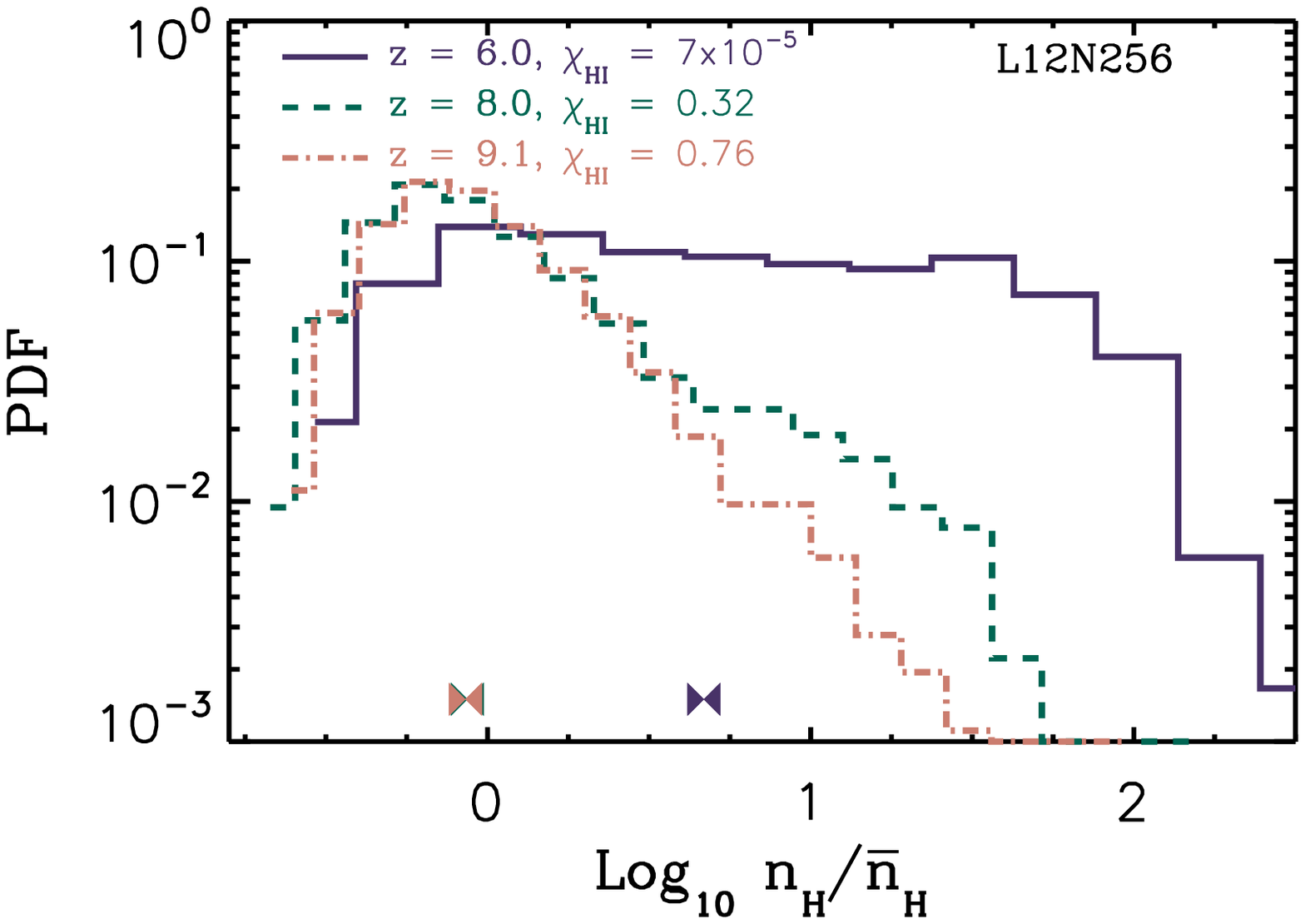}}
             \hbox{\includegraphics[width=0.5\textwidth]
             {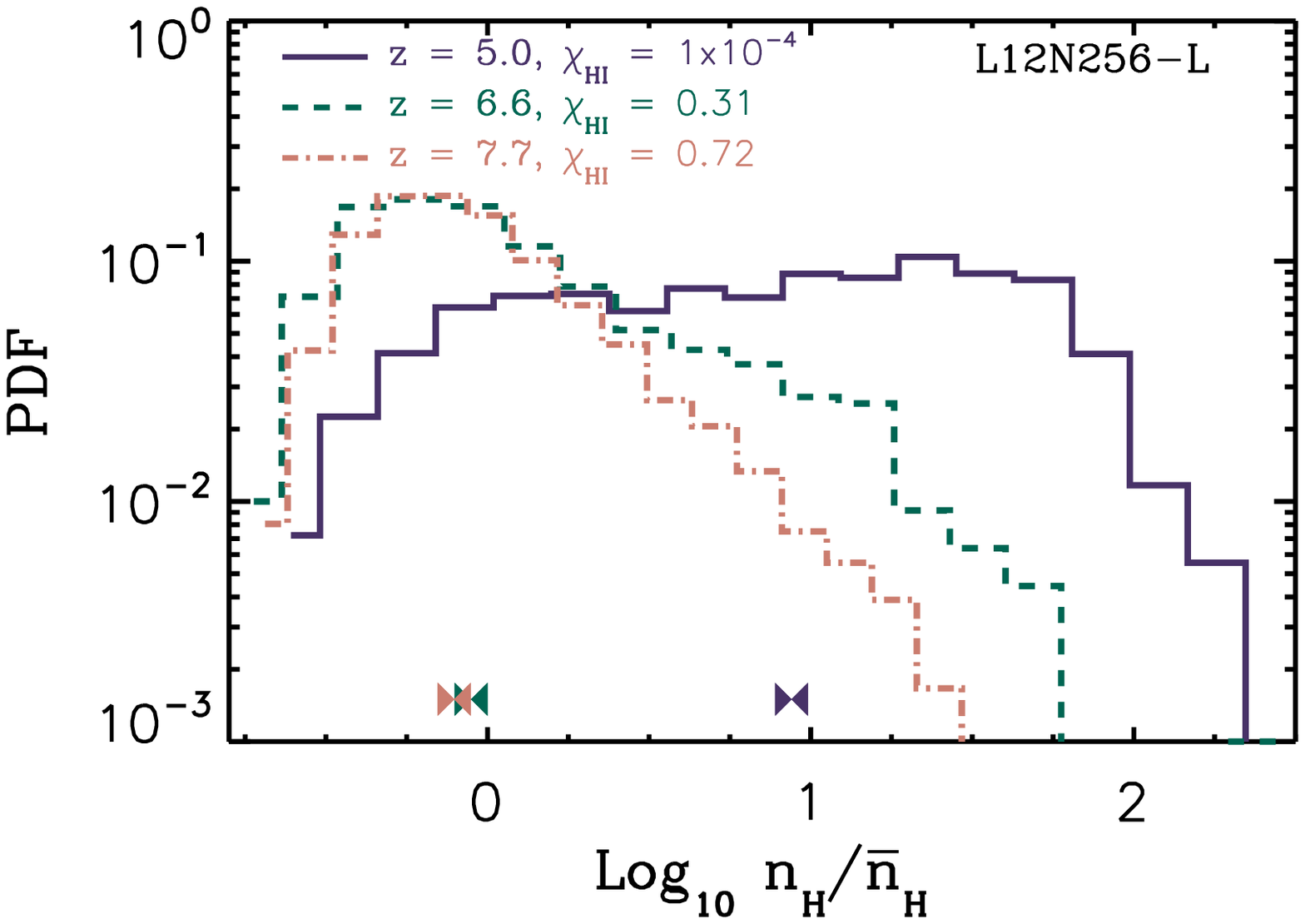}}}
\caption{Probability distribution functions of hydrogen neutral fractions (top) and hydrogen overdensities (bottom) at points along the random LOSs where the cumulative optical depth of unity (ODU points) is reached for $912 \AA$ photons. Left and right panels show the results for the \emph{L12N256} and \emph{L12N256-L} simulations. In each panel, the distribution is shown at three different times. The medians of the distributions are indicated by the symbols at the bottom of each plot, using the color of the corresponding distribution. The 3 different redshifts in the left and right panels are chosen to correspond to similar values of $\chi_{\rm{HI}}$, resulting in similar MFPs and FP distributions.}
\label{fig:HI-nH-dists}
\end{figure*}

\subsection{Distribution of free paths}
\label{sec:FPs}
While the MFP is a useful quantity for probing the evolution of the IGM during the EoR, there is more information in the full distribution of FPs along different LOS at a given time. The distribution of FPs for two different simulations with different reionization histories are shown in Fig. \ref{fig:FP-dists}. The different times shown for \emph{L12N256} and \emph{L12N256-L}, which reach a neutral fraction of 0.5 at $z \approx 8.2$ and 7.0, respectively, are chosen to roughly correspond to the same volume-weighted neutral fractions in the two simulations. The solid blue, dashed green and dot-dashed red curves correspond to volume-weighted $\HI$ fractions of $\chi_{\rm{HI}} \approx 10^{-4}$, 0.3 and 0.7, respectively. The MFP that corresponds to each redshift is indicated using the star symbol near the bottom of each plot with the color matching that of distribution function. The distributions of FPs are similar for the two simulations at similar stages of reionization, which results in similar mean free paths. Therefore, while the MFP evolves differently with time in simulations with different reionization histories, its evolution with the neutral fraction is nearly identical for the different reionization histories. 

As Fig. \ref{fig:FP-dists} shows, the shape of the FP distribution function evolves significantly during the EoR. The distribution of FPs is rather flat during the early stages of reionization but steepens after reionization is complete (e.g., at $z = 6$ and 5 for \emph{L12N256} and \emph{L12N256-L}, respectively) and shifts toward longer distances. As a result, while the MFP increases rapidly during reionization, the scatter in the FP distribution becomes smaller. This can be seen from the 15-85$\%$ distribution of the FPs shown using dotted curves around the long-dashed red curve  (which shows the MFP) in Fig. \ref{fig:MFP-evol}. 

\subsection{Physical properties of opaque regions}
\label{sec:ODUs}

For a better understanding of the connection between the physical conditions of the IGM and the MFP, it is useful to investigate the gas properties at the location where the optical depth of unity (ODU point) is reached for $912 \AA$ photons along a given LOS. Figure \ref{fig:HI-nH-dists} shows the distribution of $\HI$ fractions (top row) and gas overdensities (bottom row) of the ODU points for the \emph{L12N256} (left column) and \emph{L12N256-L} (right column) simulations at different stages of the EoR. For calculating the gas overdensity, we divide gas hydrogen number density, $n_{\rm{H}}$ by mean hydrogen number density, $\overline{n_{\rm{H}}} = \rho_{\rm{crit}}~ \Omega_{\rm{b}} ~X~ / m_{\rm{H}}$, where $\rho_{\rm{crit}}$ is the critical density of the Universe at a given redshift, $\Omega_{\rm{b}} = 0.0448$ is the baryon density parameter we use in our simulations, $X = 0.75$ is the primordial hydrogen mass fraction and $m_{\rm{H}}$ is hydrogen mass. The solid blue, dashed green and dot-dashed red curves correspond to volume-weighted $\HI$ fractions of $\chi_{\rm{HI}} \approx 10^{-4}$, 0.3 and 0.7 in both simulations. Comparison between the left and right columns shows that the ODU $\HI$ fractions and gas overdensities in the two simulations have very similar distributions at fixed IGM $\HI$ fractions.

\begin{figure*}
\centerline{\hbox{\includegraphics[width=0.5\textwidth]
             {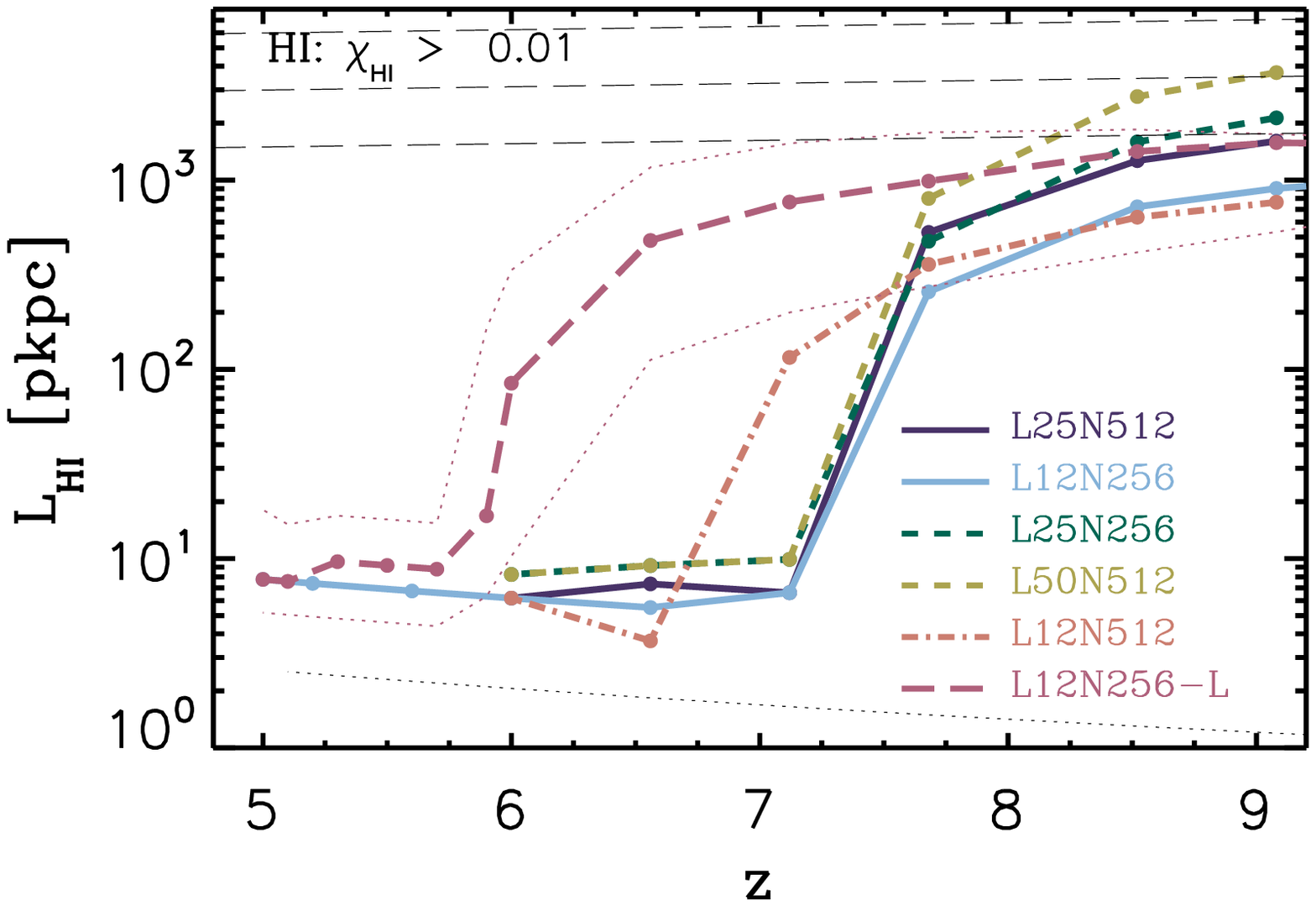}}
             \hbox{\includegraphics[width=0.5\textwidth]
             {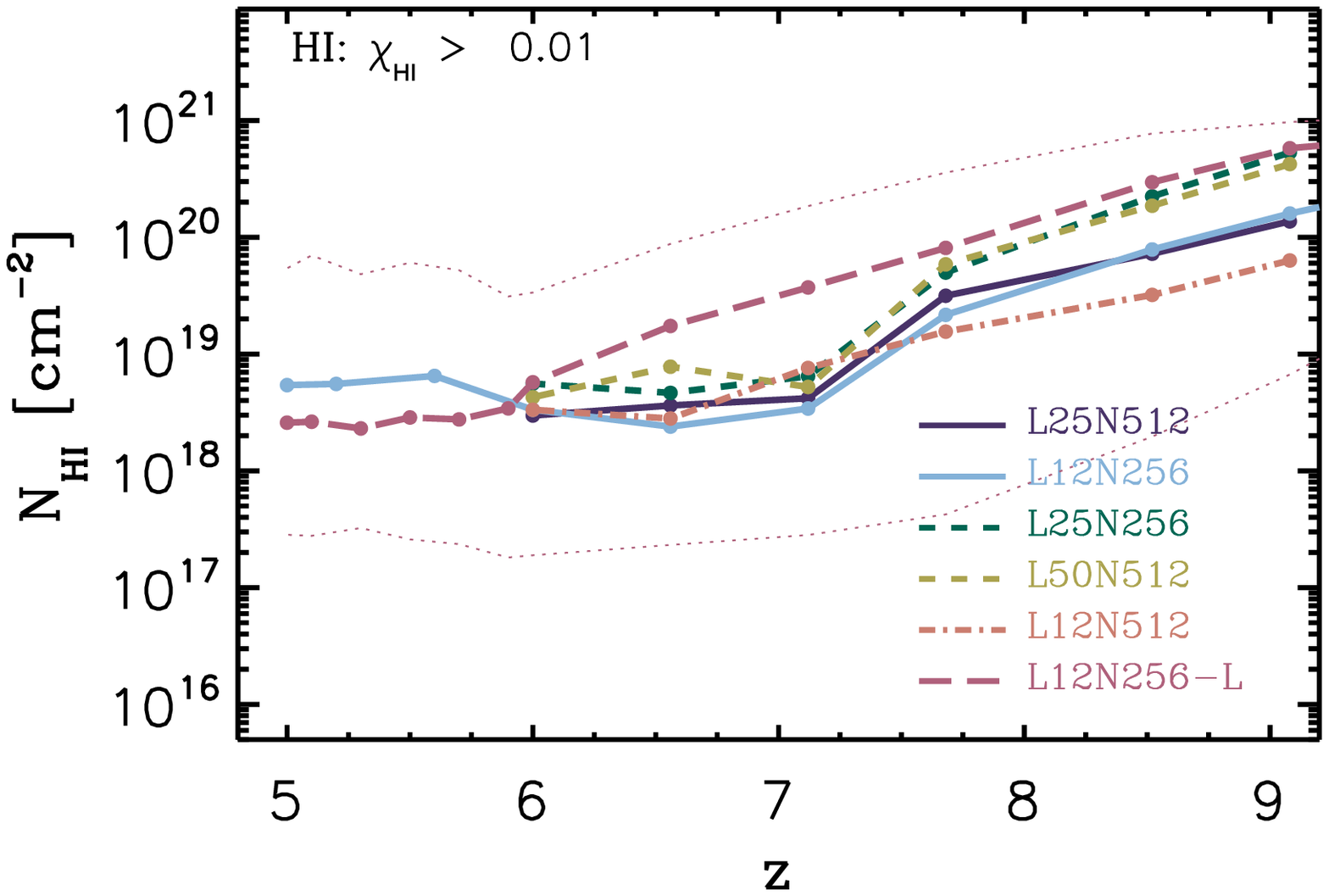}}}
\caption{The evolution of the median size (left panel) and \HI column density (right panel) of \HI regions in different Aurora simulations. \HI regions are defined as patches of gas with \HI fractions above $0.01$. The 15th-85th percentiles of the size and column density distributions of the $\HI$ regions for the \emph{L12N256-L} simulation are shown with the red dotted curves. The nearly horizontal long-dashed lines near the top of the left panel show the three different box sizes of our simulations and the dotted line near the bottom of the left panel shows the spatial resolution used for calculating the sizes, which is worse than the sub-pkpc spatial resolution of the simulations. Ater decreasing by nearly 2 orders of magnitude during reionization, the sizes converge to a typical value of $\rm{L_{HI}} \sim 10~\rm{pkpc}$. The corresponding $\HI$ column densities converge to $\NHI \sim  10^{18-19}~\cmsq$ after reionization.} 
\label{fig:LHI-evol}
\end{figure*}
\begin{figure*}
\centerline{\hbox{\includegraphics[width=0.5\textwidth]
             {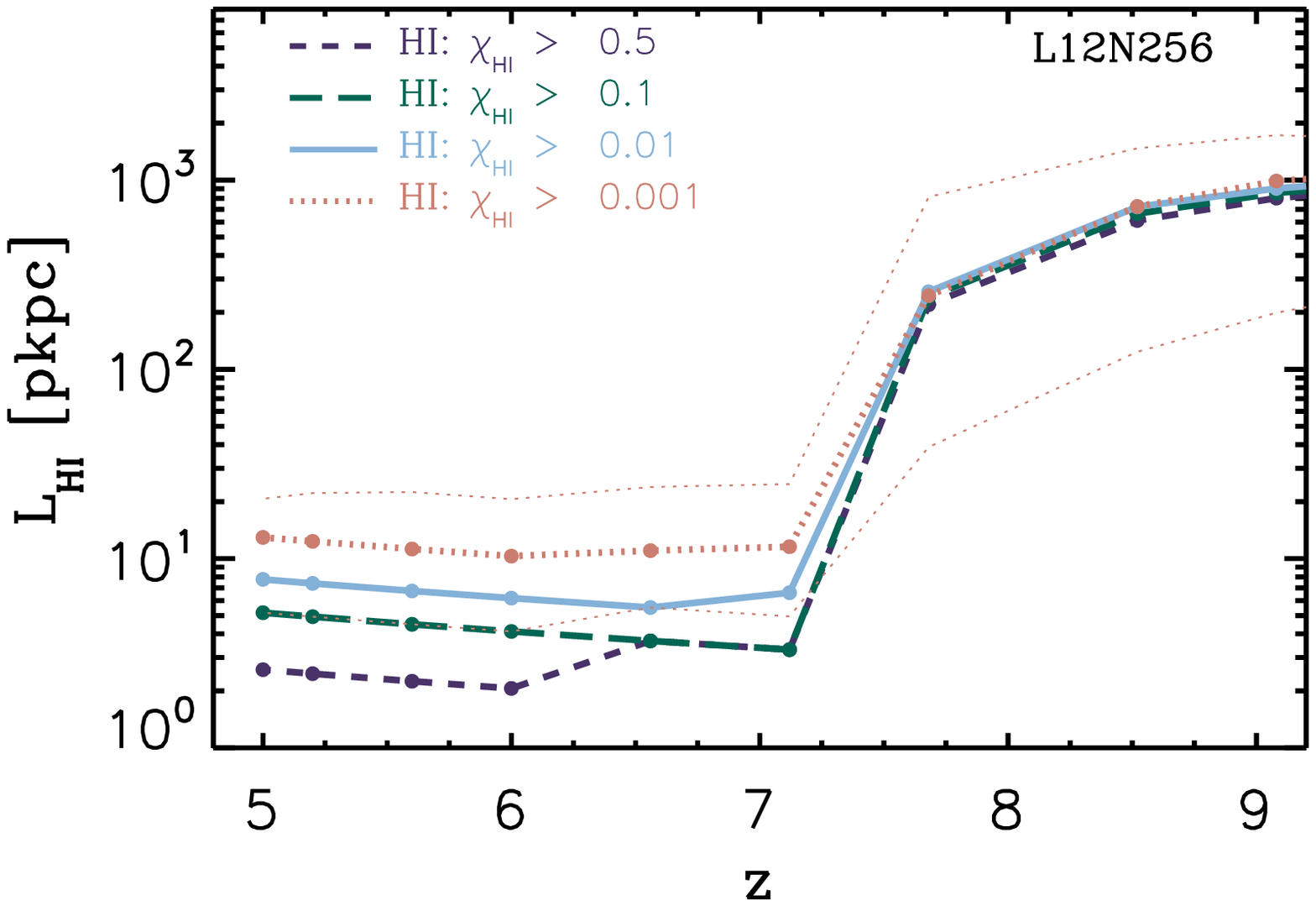}}
             \hbox{\includegraphics[width=0.5\textwidth]
             {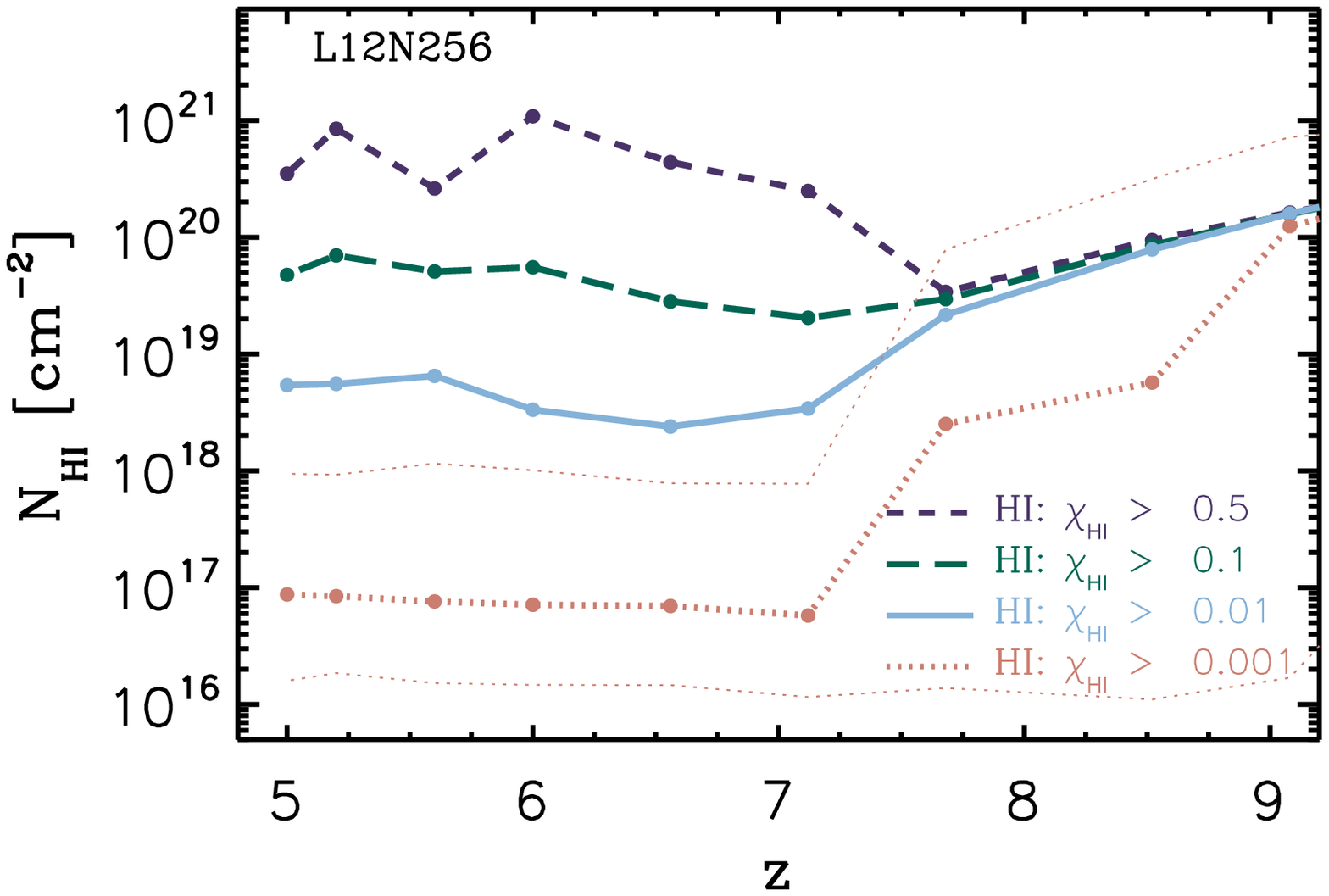}}}
\caption{The evolution of the median size of \HI regions in the \emph{L12N256} simulation (left) and their median $\HI$ column densities (right). Different curves show different \HI fraction thresholds for defining the $\HI$ regions, namely patches of gas with \HI fractions above $0.5,~0.1,~0.01$ and 0.001 (shown with blue dashed, green long-dashed, blue solid and red dotted curves, respectively). The 15th-85th percentiles of the size and column density distributions of the $\HI$ regions using the $\HI$ fraction threshold of 0.001 are shown with the thin dotted curves. Before reionization ($z \gtrsim 8$), the \HI sizes are nearly independent of the $\HI$ fraction threshold used for defining them. After reionization, the $\HI$ regions become smaller and their typical $\HI$ column density increases as their neutral fraction increases.} 
\label{fig:LHI-evol-def}
\end{figure*}

The distribution of $\HI$ fractions at the ODU points during the earlier stages of reionization (the red dot-dashed and green dashed curves in the top row of Fig. \ref{fig:HI-nH-dists}) has a strong peak at an $\HI$ fraction of unity with an extended and relatively flat tail toward lower $\HI$ fractions. As reionization proceeds, the amplitude of the dominant peak at $\chi_{\rm{HI}} \approx 1$ decreases and the tail extends to lower $\HI$ fractions. This causes the median $\HI$ fraction of the ODU points, which is unity when the mean $\HI$ fraction is $\approx 0.7$, to shift to $0.1$ by the time the mean $\HI$ fraction has decreased to $\approx 0.3$. In the early stages of reionization, when the volume-filling factor of ionized gas is relatively low, only a small number of LOS go through the ionized regions that result in larger FPs and lower ODU point $\HI$ fractions (which probe where the LOS is entering the neutral gas after passing through the ionized regions). After reionization, the distribution of the ODU point $\HI$ fractions becomes positively skewed and peaks at values comparable to, but slightly larger than the mean $\HI$ fraction. 

The distributions of ODU point densities (bottom panels of Fig. \ref{fig:HI-nH-dists}) are consistent with the general picture we discussed above. During the early stages of reionization, the ODU point density distribution peaks at around the mean hydrogen number density of the Universe (which is $\overline{\nH} \approx 2 \times 10^{-4}$, $10^{-4}$ and $4 \times 10^{-5}~\cmcb$ at $z = 9$, 7 and 5, respectively). After the completion of reionization, however, the ODU point density distribution is rather flat and its median is much larger than the mean hydrogen number density. This shows a fundamental shift in the origin of the MFP: while the mean absorption by the IGM determines the MFP at early stages and before the completion of reionization, the MFP is set by the abundance of overdense and relatively neutral (though highly ionized) systems after its completion.

\subsection{Sizse of HI regions}
\label{sec:HIsizes}

While studying different physical properties of the ODU points is useful, the power of such an analysis is limited because ODU points do not necessarily represent the properties of the gas over structures that set the MFP and its evolution. For instance, in a highly ionized universe the ODU point most likely corresponds to the outer boundary of a mostly neutral region. To gain more insight into the sizes and column densities of $\HI$ regions during reionization, we investigate the properties of the $\HI$ regions which can be identified along our LOSs as one dimensional islands of neutral hydrogen. 

During the late stages of reionization, the transition from highly ionized to neutral self-shielded gas is typically a steep function of density, which for typical photoionization rates in the IGM after reionization happens at $\nH \sim 1^{-2}$ where the $\HI$ fraction is $\chi_{\rm{HI}} \sim 0.01$ \citep{Rahmati13a,Chardin17}. Therefore, using an $\HI$ fraction threshold of $0.01$ for selecting 1D islands of clustered neutral hydrogen along the LOS is reasonable. The evolution in the median proper size of the $\HI$ regions defined in this manner is shown in the left panel of Figure \ref{fig:LHI-evol}. Curves with different line styles and colors show different Aurora simulations. To illustrate the typical scatter in the distribution of the $\HI$ sizes, the 15th-85th percentiles are shown for the \emph{L12N256-L} simulation using the red dotted curves. For all the simulations, the \HI sizes evolve qualitatively similarly to the IGM \HI fraction. After decreasing by 2 orders of magnitude during reionization, the $\HI$ sizes converge to a typical value of $\rm{L_{HI}} \sim 10~\rm{pkpc}$, which increases very slowly after the completion of reionization (i.e., at $z \lesssim 6$ for the \emph{L12N256-L} simulation and at $z \lesssim 7$ for the other simulations). The similarity of all the simulations with similar reionization histories, and the fact that the sizes of $\HI$ regions are much larger than the sub-kpc resolution of the simulations, suggests that all the simulations resolve the main $\HI$ absorbers in the IGM. The \emph{L12N256-L} simulation with late reionization shows a trend that is very similar to the rest of the simulations, but with a fixed negative shift along the redshift axis.

The right panel of Figure \ref{fig:LHI-evol} shows the evolution of the median $\HI$ column density of the $\HI$ regions identified as neutral patches with $\chi_{\rm{HI}} > 0.01$. The evolution of the $\HI$ column densities closely follows that of the size of the $\HI$ regions shown in the left panel: it decreases by 1-2 orders of magnitude before reaching an asymptotic and slowly varying value of $\NHI \sim 10^{18.5}~\cmsq$ at $z \lesssim 7$. The 15th-85th percentile scatter in the distribution of $\HI$ column densities at a given redshift and for each simulation (shown with red-dotted curves for the \emph{L12N256-L} simulation) is, however, much larger than for the $\HI$ sizes. This can be understood by noting that the scatter in the $\HI$ column density results from the scatter in the sizes, densities and neutral fractions of the $\HI$ regions.

Figure \ref{fig:LHI-evol-def} shows how the results we discussed above change by varying the neutral hydrogen fraction threshold used for identifying the $\HI$ regions in the \emph{L12N256} simulation. As the left panel shows, before the completion of reionization, different $\HI$ fraction thresholds result in nearly identical sizes, which all decline very rapidly during reionization (at $z \approx 7-8$). At later stages, on the other hand, the typical size of the $\HI$ regions does not evolve strongly, but shrinks if we increase the $\HI$ fraction threshold used to identify them. The change in the typical size is generally much smaller than the change in the $\HI$ fraction threshold. This indicates a rather sharp transition in the $\HI$ fraction of predominantly neutral systems, which, as discussed above, is expected if those systems self-shield against a background radiation field \citep[e.g.,][]{Rahmati13a,Rahmati13b}. 

The $\HI$ column densities corresponding to the different $\HI$ fraction thresholds used in the left panel of Figure \ref{fig:LHI-evol-def} are shown in its right panel. Contrary to the sizes, after reionization the difference in the final $\HI$ column densities of the $\HI$ regions is similar to or larger than the difference between the $\HI$ fraction used to identify them. This is because the $\HI$ column density of a given system is the product of its size, its neutral fraction and its density. For the same reason, and as mentioned above, the scatter in the $\HI$ column density is also larger than the scatter in the $\HI$ sizes. 

It is worth noting that after reionization, the typical size of the $\HI$ systems presented in Figure \ref{fig:LHI-evol-def} closely follows the local Jeans scale, indicating that the systems are close to local hydrostatic equilibrium \citep{Schaye01}. As a result one can use the Jeans length to relate the typical size, column-density and density of absorbers. Following \citet{Schaye01}, the Jeans length can be expressed as a function of the gas density:
\begin{equation}
L_{\rm{J}}  \sim 1.3 ~{\rm{kpc}} ~ f^{\frac{1}{2}}  \left(\frac{\nH}{1 \cmcb}\right)^{-\frac{1}{2}}  \left(\frac{T}{10^4~{\rm{K}}}\right)^{\frac{1}{2}} \left(\frac{\mu}{0.59}\right)^{-\frac{1}{2}},
\label{eq:Ljeans}
\end{equation}
for an ideal, monoatomic gas with density $\nH$, temperature $T$, adiabatic index $\gamma = 5/3$ and hydrogen mass fraction of $X = 0.75$. Here, $f = f_{\rm{g}} / f_{\rm{th}}$ is the ratio between the gas fraction in the medium, $f_{\rm{g}}$, and $f_{\rm{th}}$ which is the ratio between the thermal and total pressure of the system \citep{Schaye08}. The mean particle mass of $\mu = 0.59$ is appropriate for ionized gas while $\mu = 1.23$ should be used for neutral gas (for a hydrogen mass fraction of $X = 0.75$). Assuming the gas pressure is fully thermal, using the cosmic baryon fraction for the gas fraction (i.e., $f_{\rm{g}} = \Ob / \Om = 0.17$), $\mu = 0.59$ and using $T = 10^4$ K, equation \eqref{eq:Ljeans} becomes:
\begin{equation}
L_{\rm{J}}  \sim 0.5 ~{\rm{kpc}}  \left(\frac{\nH}{1 \cmcb}\right)^{-\frac{1}{2}}.
\label{eq:Ljeans2}
\end{equation}

The typical densities corresponding to each $\HI$ fraction in the \emph{L12N256} simulation at $z\approx 6$ are $\nH \sim 0.01$, $0.003$ \& $0.001~\cmcb$ for $\HI$ fractions $0.1$, $0.01$ and $0.001$, respectively. Using equation \eqref{eq:Ljeans2}, those typical densities correspond to Jeans lengths of $L_{\rm{J}} \sim 5$, $9$ \& $16$ pkpc, respectively. As the left panel of Figure \ref{fig:LHI-evol-def} shows, this matches the simulation results very well. Note, however, that at earlier times, e.g., $z > 7$ for the \emph{L12N256} simulation, the relation between $\HI$ fraction and density found by imposing a certain neutral fraction threshold becomes somewhat ill-defined and makes their sizes deviate from what is expected based on their local Jeans lengths. As mentioned above, at those earlier times the typical sizes of the $\HI$ regions increases sharply with redshift and is insensitive to the $\HI$ fraction threshold.

\begin{figure*}
\centerline{\hbox{\includegraphics[width=0.5\textwidth]
             {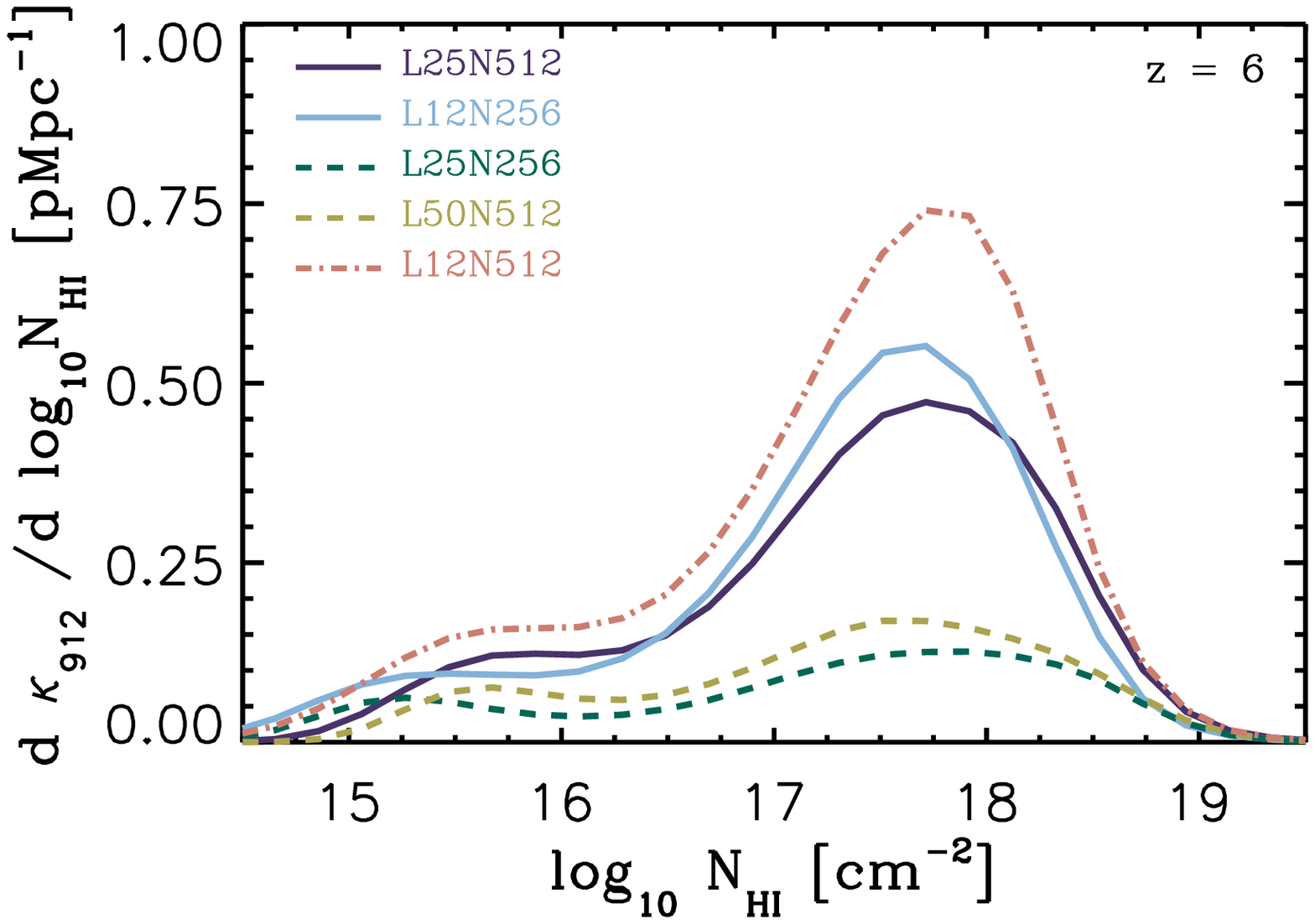}}
             \hbox{\includegraphics[width=0.5\textwidth]
             {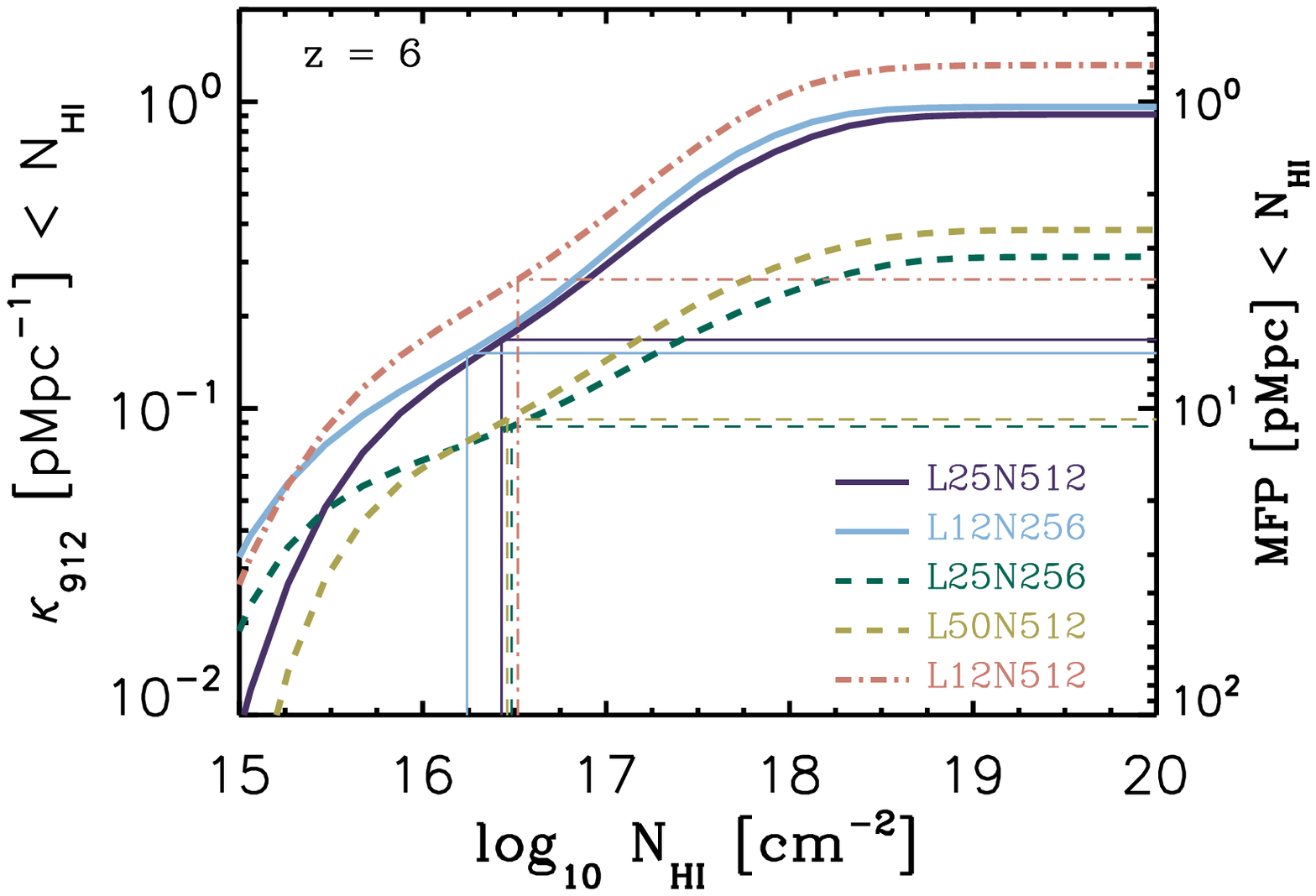}}}
\caption{The left panel shows the differential opacity (per proper Mpc; see equation \eqref{eq:opa-diff}), as a function of $\HI$ column density for different simulations at $z = 6$. The right panel shows the cumulative opacity (per unit proper Mpc; see equation \eqref{eq:opa}) resulting from $\HI$ column densities lower than the value plotted along the x-axis. The right axis in the right panel shows the corresponding MFP (i.e., $1/ \kappa_{912}$) while the actual, directly measured MFP of each simulation is shown using a horizontal line with the corresponding color and style. While the opacity distributions peak at $\NHI \sim 10^{18}~\cmsq$, absorbers with $\NHI \lesssim 10^{16.5}~\cmsq$ are the main contributors to the mean absorption of hydrogen ionizing photons.}
\label{fig:Kappa}
\end{figure*}
\begin{figure*}
\centerline{\hbox{\includegraphics[width=0.5\textwidth]
             {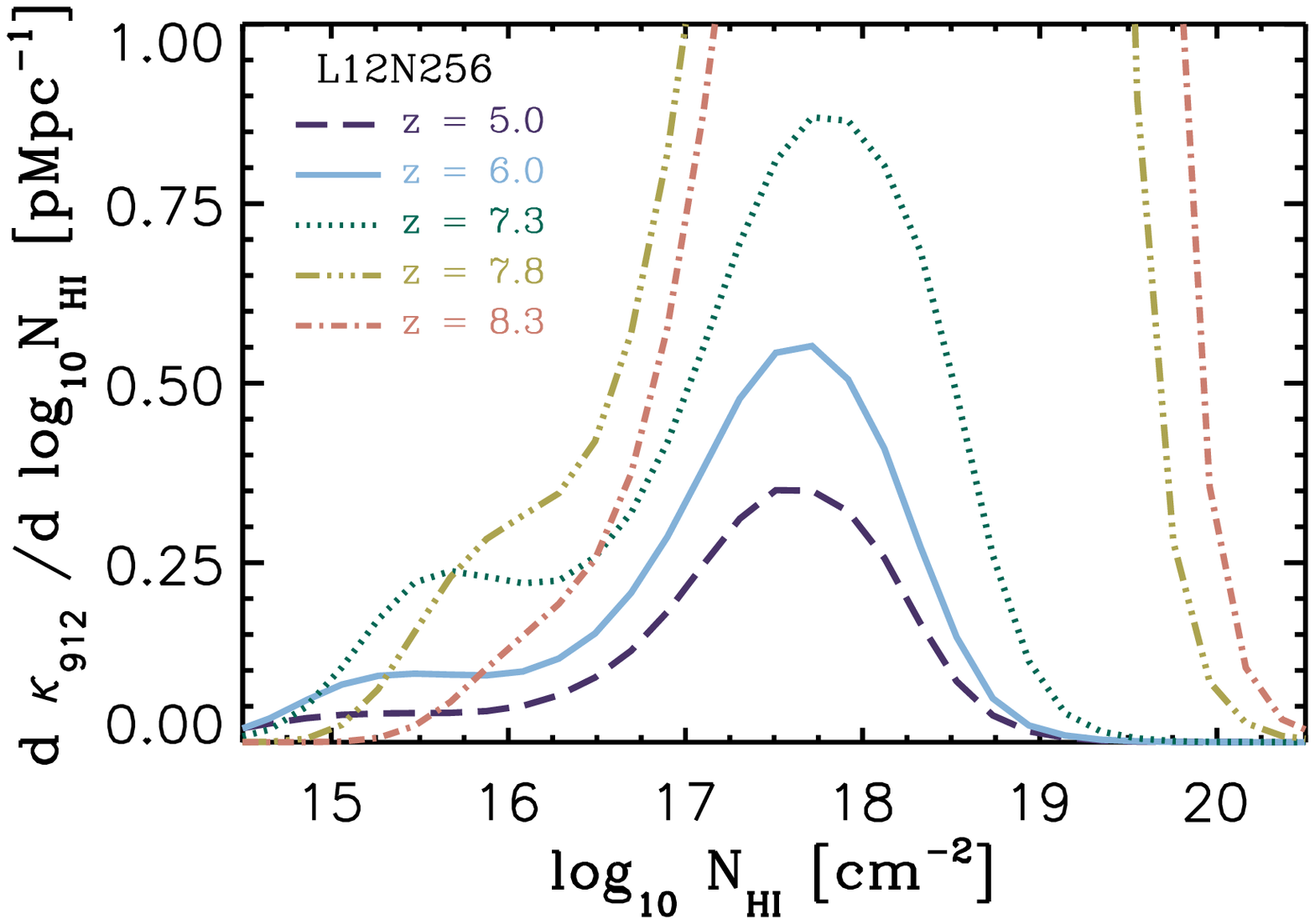}}
             \hbox{\includegraphics[width=0.5\textwidth]
             {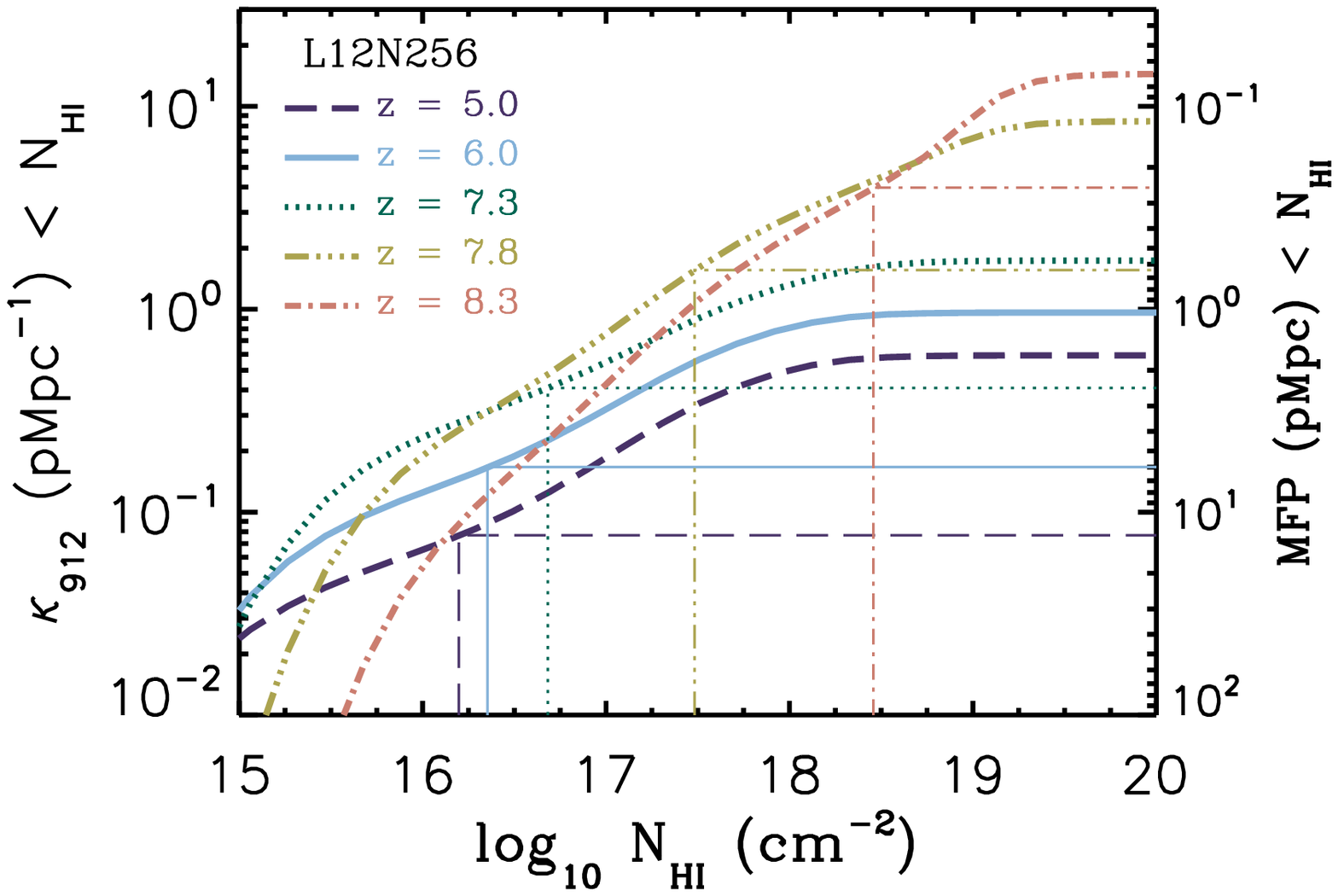}}}
\caption{The left panel shows the differential opacity (per proper Mpc; see equation \eqref{eq:opa-diff}), as a function of $\HI$ column density for the \emph{L12N256} simulation at different redshifts. The right panel shows the cumulative opacity (per unit proper Mpc; see equation \eqref{eq:opa}) resulting from $\HI$ column densities lower than the value plotted along the x-axis. The right axis in the right panel shows the corresponding MFP (i.e., $1/ \kappa_{912}$) while the actual, directly measured MFP at each redshift is shown using a horizontal line with the corresponding color and style. The opacity distributions always peak at $\NHI \gtrsim 10^{17.5}~\cmsq$ but the typical $\HI$ column density of absorbers which set the mean free path of hydrogen ionizing photons decreases rapidly at earliy times before converging to $\NHI \sim 10^{16.5}~\cmsq$ after reionization.}
\label{fig:KappaEvol}
\end{figure*}

\subsection{The origin of the MFP}
\label{sec:MFPorigin}
The contribution of different $\HI$ column densities to the absorption of 1 Ry photons depends on the $\HI$ distribution. The $\HI$ column density distribution, $\frac{d\mathcal{N}}{d\NHI dr} \left(\NHI\right) d\NHI$, can be defined as the number of absorbers with a column density between $\NHI$ and $\NHI + d\NHI$, per unit proper distance, $dr$. We can further define the opacity for 1 Ry photons, $\kappa_{912}$, as the rate of change in the optical depth at $912~\AA$, $\tau_{912}$, per unit proper distance:
\begin{equation}
\kappa_{912} \equiv \frac{d\tau_{912}}{dr}.
\label{eq:opa}
\end{equation}
This yields \citep[e.g.,][]{Meiksin93,Prochaska09, Rudie13}:
\begin{equation}
\kappa_{912} = \int^{\infty}_{0} \frac{d\mathcal{N}}{d\NHI dr} \left(\NHI\right) (1 - e^{-\NHI \sigma_{912}})~d\NHI.
\label{eq:opa1}
\end{equation}

To calculate $\kappa_{912}$ in our simulations, we first compute the $\HI$ column density distribution function by calculating the projected column densities on uniform 2-dimensional grids with $2500^2$ and $5000^2$ cells for simulations with initial SPH particle numbers of $256^3$ and $512^3$, respectively \citep[see][]{Rahmati13a, Rahmati15, Rahmati16}. Furthermore, to prevent the overlap of distinct systems along the LOS, we divide the full box into 32 slices along the projection direction. Then we calculate the differential opacity, $\frac{d\kappa_{912}}{d\NHI} \left(\NHI\right)$, by choosing a narrow range of $\HI$ column densities, $\NHI < N_{\rm{HI}, i} < \NHI + \Delta \NHI$, and summing over the contribution of all systems within that range of column densities \citep[e.g.,][]{Rudie13}: 

\begin{eqnarray}
&{}&\frac{d\kappa_{912}}{d\log{\NHI}} \left(\NHI\right) = \frac{1}{\Delta \log{\NHI}} \times \nonumber \\
&{}& \int^{\NHI + \Delta \NHI}_{\NHI} \frac{d\mathcal{N}}{d\NHI dr} \left(\NHI\right) (1 - e^{-\NHI \sigma_{912}})~d\NHI.
\label{eq:opa-diff}
\end{eqnarray}
The left panel of Figure \ref{fig:Kappa} shows the differential opacity, $\frac{d\kappa_{912}}{d\NHI} \left(\NHI\right)$, versus $\HI$ column density for our different simulations at $z = 6$ and using $\HI$ column density bins of $\Delta \log_{10}{\NHI} = 0.2$. The opacity distributions all peak at $\NHI \sim 10^{17-18}\cmsq$ due to the sharp increase of $1 - \exp{(-\NHI \sigma_{912})}$ in this column density range, i.e., the fraction of absorbed photons, combined with the steep decline in the abundance of absorbers with higher $\HI$ column densities. The location of the peak in the opacity distributions suggests that Lyman Limit systems determine the MFP, as has been suggested for lower redshifts \citep[e.g.,][]{Rudie13}. However, this is only true if the cumulative opacity of the more frequent but less opaque absorbers with lower $\HI$ column densities is not high enough. Indeed, as the right panel of Figure \ref{fig:Kappa} shows, for all our simulations the cumulative opacity reaches the value corresponding to our directly calculated MFP (see $\S$\ref{sec:MFP}) at $\NHI \sim 10^{16.5}\cmsq$ (see the shaded vertical bar). Moreover, the differences between the MFPs in the different simulations are almost identical to the ratios between their cumulative opacities at $\NHI \approx 10^{16.5}\cmsq$. This suggests that shortly after the completion of reionization, the abundance of absorbers with $\NHI \approx 10^{16.5}\cms$ determinines the MFP. 

The evolution of the differential (left) and cumulative (right) opacities for the \emph{L12N256} simulation are shown in Figure \ref{fig:KappaEvol}. The differential opacity distribution evolves significantly, particularly during the early stages of reionization ($z > 7$). However, the distribution always peaks at $\NHI \gtrsim 10^{17-18}\cmsq$. The $\HI$ column density at which the cumulative opacity corresponds to the directly calculated MFP, on the other hand,  continuously shifts towards lower values as reionization proceeds. While at $z = 8.2$ the dominant $\HI$ column densities for the MFP are $\NHI \approx 10^{18.5} \cmsq$, after reionization and for $z \lesssim 7$, the relevant column densities are $\NHI \sim 10^{16-17} \cmsq$. We note that at lower redshifts ($z < 5$) the combined evolution of the $\HI$ column density distribution function and the MFP, shifts the column densities at which the cumulative opacity corresponds to the MFP towards higher values. Using the $\HI$ column density distribution functions calculated for the EAGLE simulations from \citet{Rahmati15}, together with observational constraints on the value of the MFP at $z = 3-5$, we found that the MFP at those redshifts is set by absorbers with $\NHI \gtrsim 10^{18} \cmsq$, as suggested by previous observational studies \citep[e.g.,][]{Rudie13}.

\begin{figure*}
\centerline{\hbox{\includegraphics[width=0.5\textwidth]
             {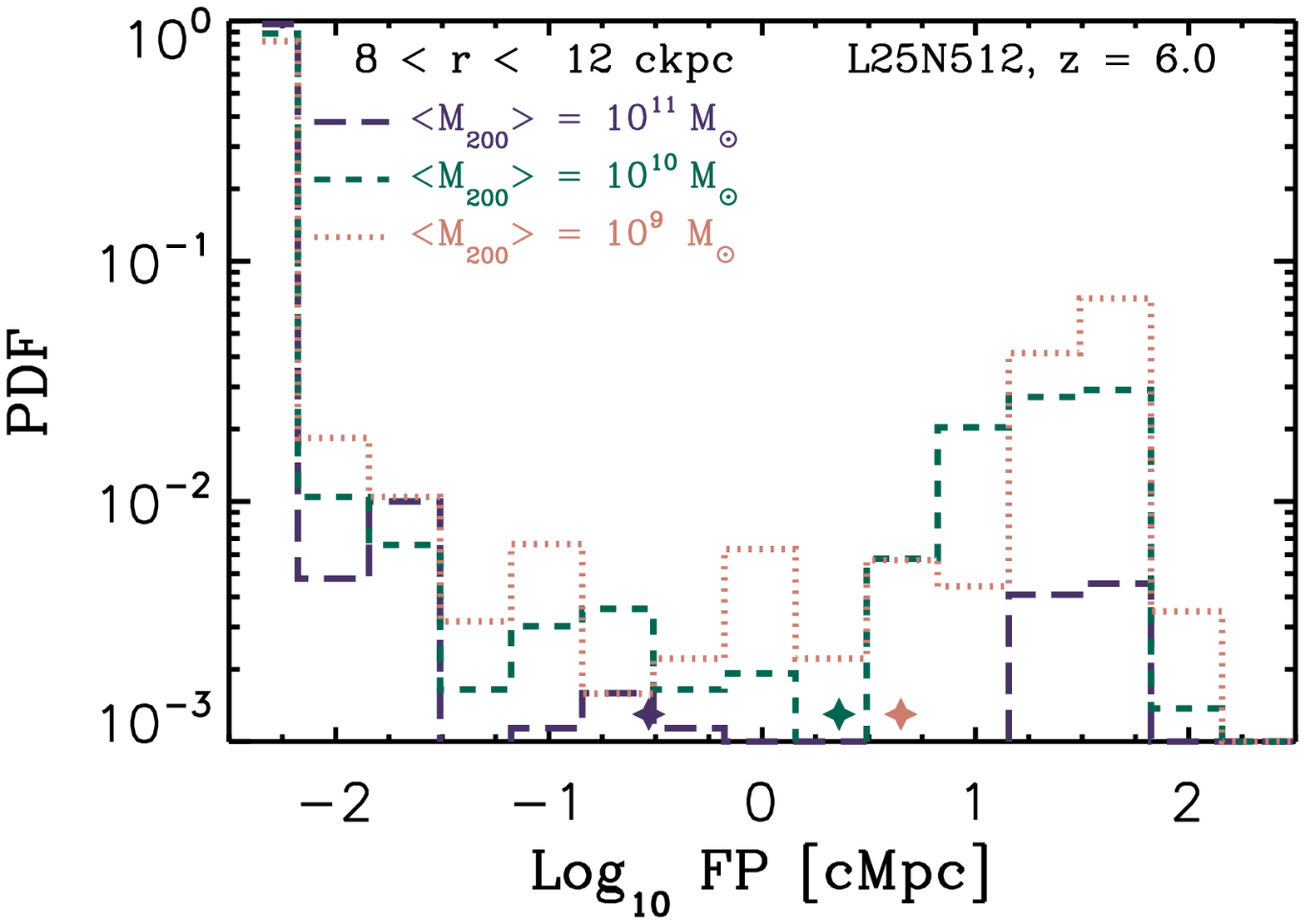}}
             \hbox{\includegraphics[width=0.5\textwidth]
             {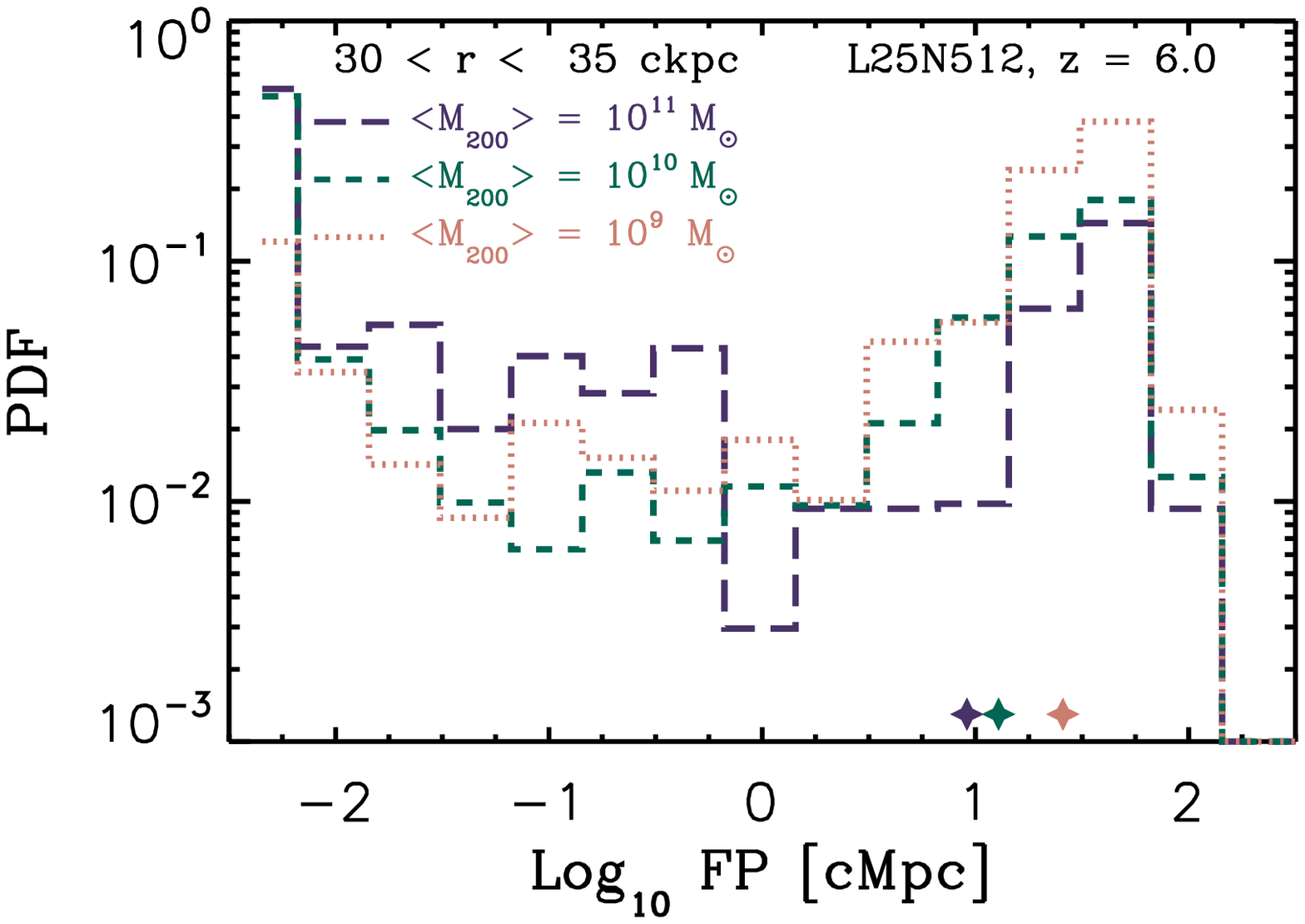}}}
\centerline{\hbox{\includegraphics[width=0.5\textwidth]
             {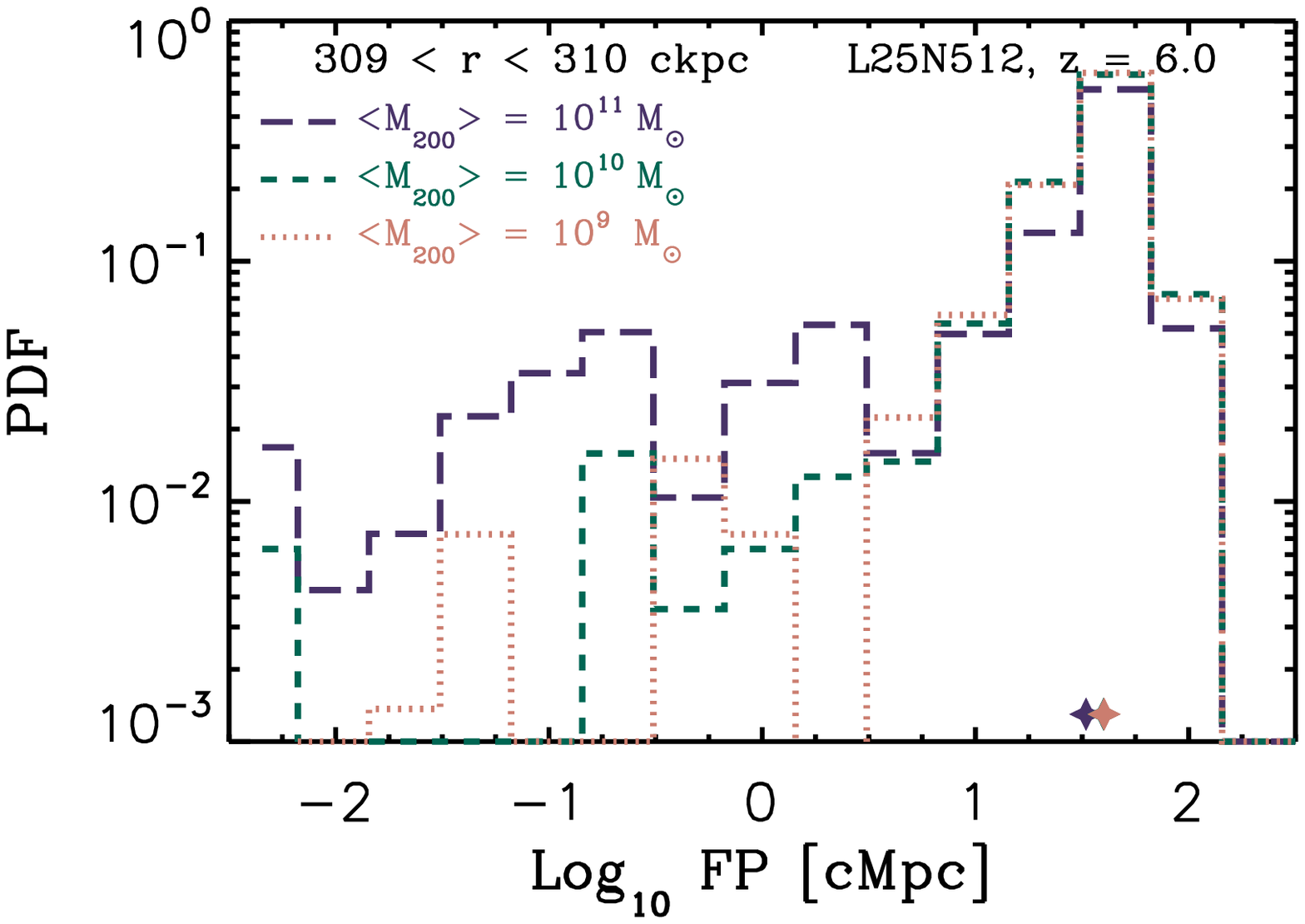}}
             \hbox{\includegraphics[width=0.5\textwidth]
             {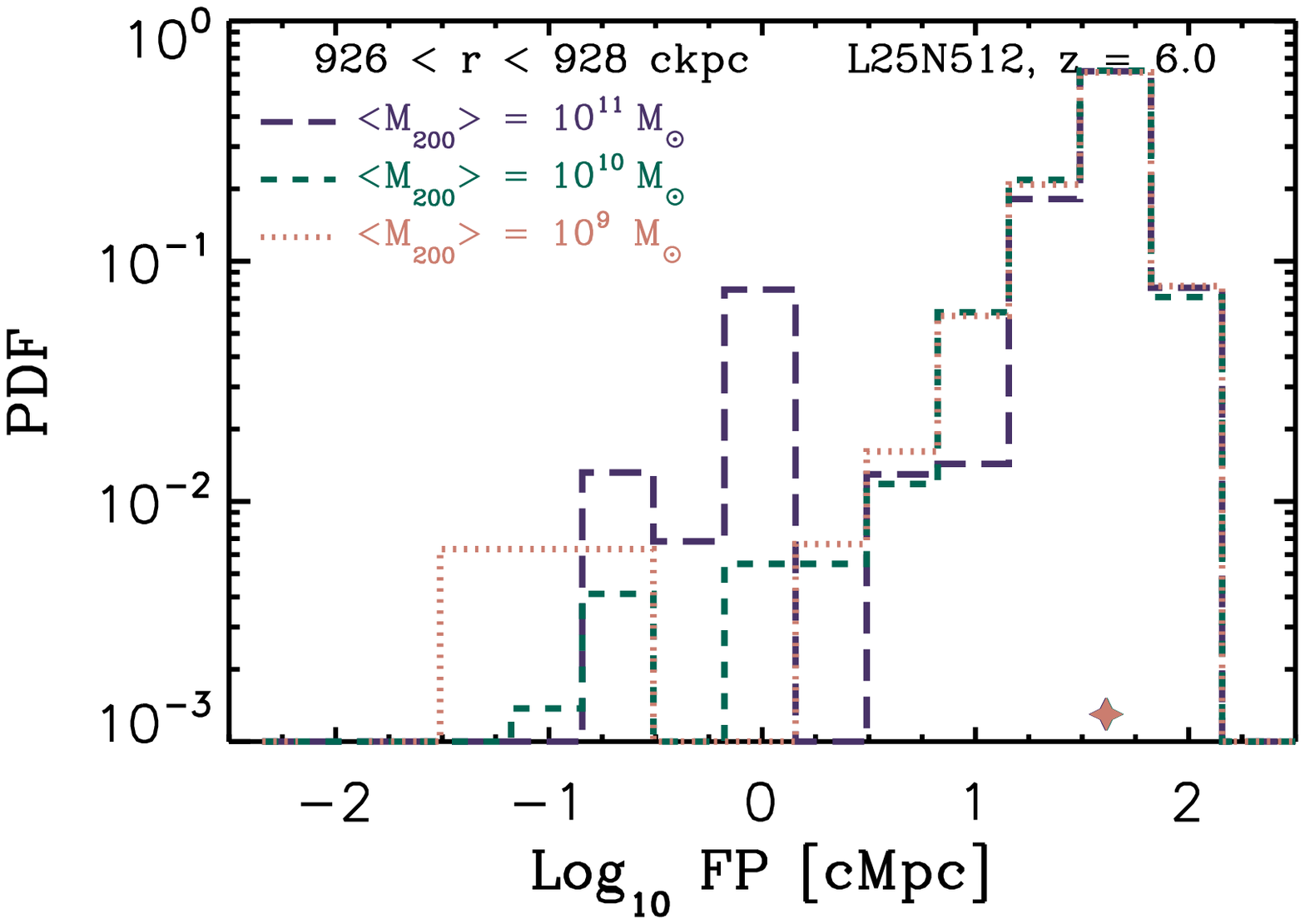}}}
\caption{Distribution of distances required for hydrogen ionizing photons to reach an optical depth of unity along lines-of-sight starting at different distances from galaxies (indicated in the top-left corner of each panel) with different halo masses in the \emph{L25N512} simulation at $z = 6$. Blue long-dashed, green dashed and red dotted curves show the distributions for halos with average masses $\left<\rm{M_{200} / M_{\odot}}\right>  = 10^{11}$, $\left< {\rm{M_{200} / M_{\odot}}}\right>  = 10^{10}$ and $\left<{\rm{M_{200} / M_{\odot}}} \right> =10^{9}$, respectively. From top-left to bottom-right the panels show the results for distances $r \approx 10$, $30$, $300$ and $1000$ ckpc. The mean value of all FPs for each distribution is shown near the bottom of each panel using star symbols with colors matching those of the histograms. The distribution of FPs is bimodal close to galaxies. The low-FP-end peak becomes weaker with increasing distance from galaxies and becomes very similar to the distribution of randomly selected FPs. The fraction of LOS with short/long FPs (e.g., FP $< 1$ cMpc / FP $> 1$ cMpc) increases with increasing halo mass. The significance of this trend, however, becomes weaker with increasing distance from galaxies.}
\label{fig:galFP-dists}
\end{figure*}
\begin{figure*}
\centerline{\hbox{\includegraphics[width=0.5\textwidth]
             {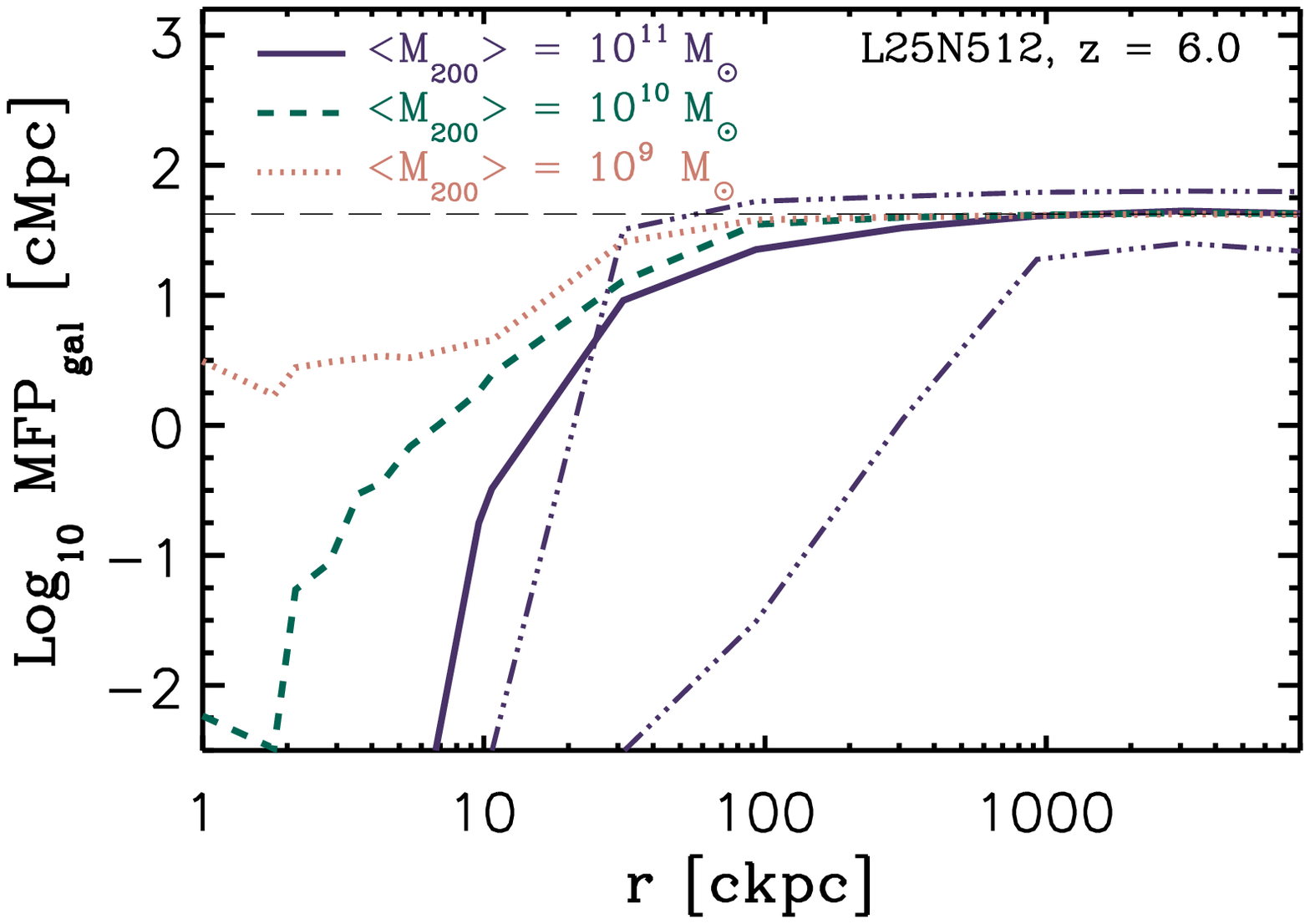}}
             \hbox{\includegraphics[width=0.5\textwidth]
             {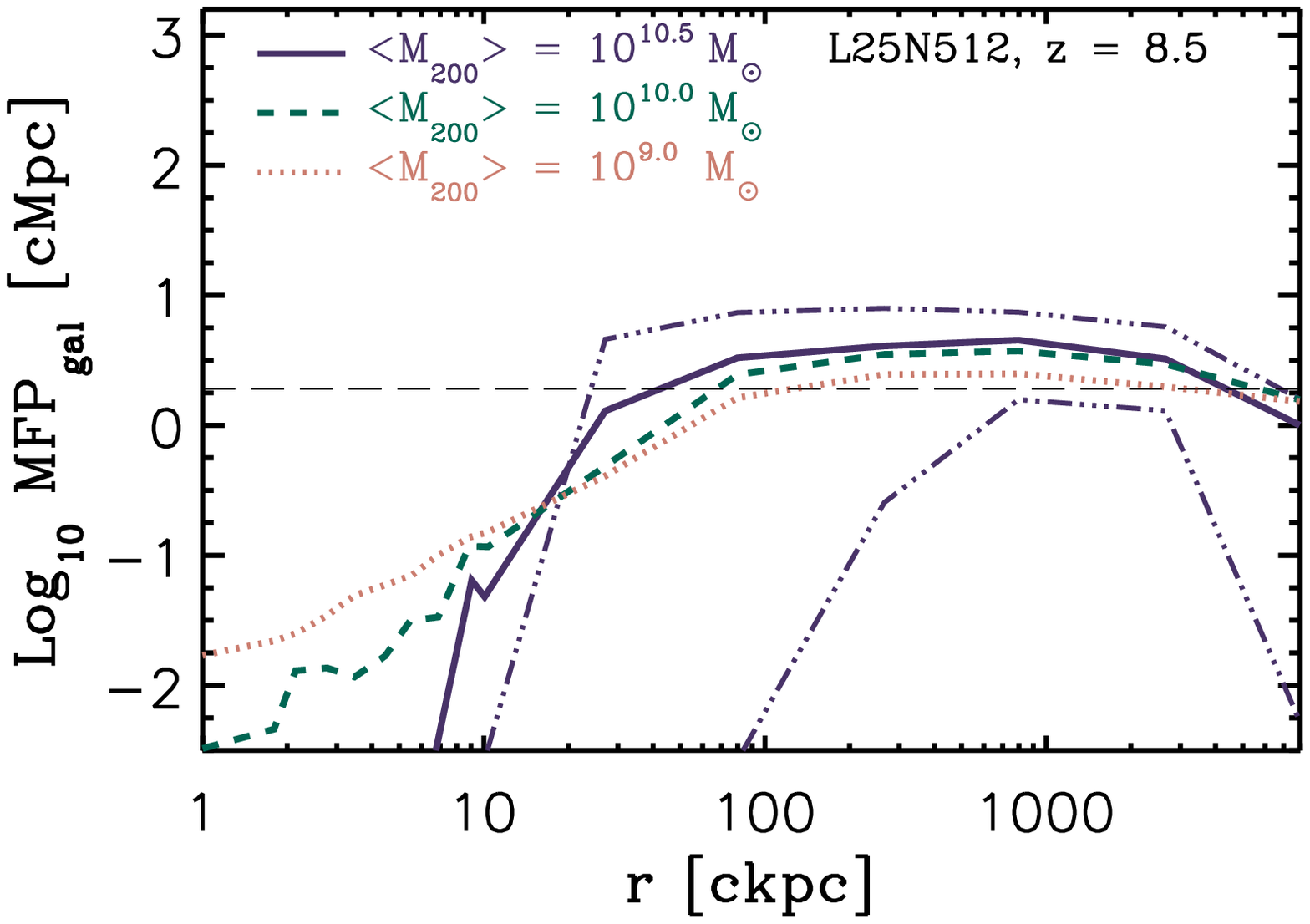}}}
\caption{The MFP measured close to galaxies, $\rm{MFP_{gal}}$ as a function of starting distance from the galaxy for different halo masses in the \emph{L25N512} simulation at $z = 6$ (left) and $z = 8.5$ (right). Blue solid, green dashed and red dotted curves show halos with average masses $\left<\rm{M_{200} / M_{\odot}}\right>  = 10^{11}$ ($\left<\rm{M_{200} / M_{\odot}}\right>  = 10^{10.5}$ for the right panel), $\left< {\rm{M_{200} / M_{\odot}}}\right>  = 10^{10}$ and $\left<{\rm{M_{200} / M_{\odot}}} \right> =10^{9}$, respectively. The 15th-85th percentiles of the FP distribution for the highest halo mass bin is shown with blue dot-dot-dashed curves. The MFP measured in a random location in the IGM is shown with the horizontal dashed curve in each panel. At both redshifts, $\rm{MFP_{gal}}$ is much smaller than the cosmic MFP for $r \lesssim 10-30$ ckpc, where the $\rm{MFP_{gal}}$ is set by the few transparent LOS while most of the LOS remain opaque (the mean is larger than the 85th percentile of the FP distribution). The $\rm{MFP_{gal}}$ close to galaxies decreases with increasing halo mass. At $z = 8.5$, when the ionized bubbles around galaxies have not yet merged, $\rm{MFP_{gal}}$ becomes longer than the IGM MFP  for $r \gtrsim 30$ ckpc. In this regime, $\rm{MFP_{gal}}$ increases with increasing halo mass. At large distances, where the starting point of the $\rm{MFP_{gal}}$ calculation drops outside of the ionized bubbles, $\rm{MFP_{gal}}$ decreases again with increasing distance (at $r \gtrsim 3$ cMpc).}
\label{fig:GalMFP}
\end{figure*}

\subsection{Reduction in the MFP close to galaxies}
\label{sec:GalMFP}

The MFP of ionizing photons, which is a measure for the opacity of the IGM, is often combined with an assumed ionizing photon escape fraction (from galaxies) to connect the total emissivity of ionizing photons in the Universe to the background radiation field which is formed due to the superposition of individual sources after reionization \citep[e.g.,][]{Miralda03,Kuhlen12,Becker15}. However, because galaxies do not form in random locations in space, the MFP for a photon emitted by a galaxy might be different from the cosmic MFP. Moreover, the radiation field close to galaxies could be enhanced and affect the abundance of $\HI$ systems close to them \citep[e.g.,][]{Rahmati13b,Becker15}. 

To quantify those effects, we calculate the MFP starting from the locations of galaxies, $\rm{MFP_{gal}}$, instead of starting from a random location, which is what we have done so far in this work to calculate the MFP. For each galaxy, we calculate several LOS passing through the vicinity of the galaxy\footnote{We select 100 LOS for the galaxies with $\rm{M_{200} / M_{\odot}}  > 10^{10.8}$ in our simulations and use 20 LOSs per galaxy for galaxies in other mass bins. We select the mass bins such that for each bin, we get $\approx 2000$ different LOSs.} and very close to its centre. . Then we calculate the FPs along each LOS as described before, but starting from a certain distance, $D$, away from the centre of the galaxy\footnote{While each LOS has a fixed randomly selected impact parameter with respect to the center of each galaxy which is always shorter than $5$ comoving kpc, by varying the starting point of the FP calculation along the LOS we can achieve a wide range of effective distances for the starting point of the FP calculation.} and integrating outwards (away from the galaxies) until we reach an optical depth of unity, as explained in section \ref{sec:MFPcalc}. 

The full distributions of FPs at different distances from galaxies are shown in Figure \ref{fig:galFP-dists} for three different halo mass bins in the \emph{L25N512} simulation at $z = 6$. Each panel in this figure shows the result for a given distance for the starting point of the FP calculation. Panels from top-left to bottom right show the results for distances $r \approx 10$, $30$, $300$ \& $900$ ckpc, respectively. In each panel, the distribution of FPs is shown for halo mass bins with average masses of $\left<\rm{M_{200} / M_{\odot}}\right>  = 10^{11}$, $\left< {\rm{M_{200} / M_{\odot}}}\right>  = 10^{10}$ and $\left<{\rm{M_{200} / M_{\odot}}} \right> =10^{9}$ with blue long-dashed, green dashed and red dot-dashed histograms, respectively\footnote{The mean stellar mass corresponding to each halo mass bin is $\left<\rm{M_{\star} / M_{\odot}}\right>  = 10^{9.4}$, $\left< {\rm{M_{\star} / M_{\odot}}}\right>  = 10^{7.8}$ and $\left<{\rm{M_{\star} / M_{\odot}}} \right> =10^{6.4}$, respectively}. The mean FP (i.e., MFP) corresponding to each histogram is shown using star symbols near the bottom of each panel. 

To first order, the FP distributions of all three mass bins change similarly with increasing distance from the galaxies. The FP distribution is bimodal close to galaxies (e.g., at $r \approx 10$ and $30$ ckpc as shown in the top tow panels of Fig. \ref{fig:galFP-dists}) with peaks at $ < 10$ ckpc and at around $\approx 30$ cMpc which corresponds to the IGM MFP at $z = 6$. Note that this bimodality in the distribution of FPs very close to galaxies (at $r \lesssim 10$ ckpc) suggests that the fraction of ionizing photons that can escape from the CGM to the IGM is mainly driven by the small number of LOS that have rather low opacities (i.e., long FPs) while the majority of LOS remain opaque \citep{Gnedin08}. Increasing the distance of LOS from galaxies increases the fraction of LOS that have FPs comparable to the IGM MFP at the expense of decreasing the fraction of opaque LOS with short FPs. At large distances from the galaxies, the distribution of FPs becomes very similar to the distribution of FPs for randomly selected LOS (compare the bottom-right panel of Figure  \ref{fig:galFP-dists} with the solid blue histogram in the left panel of Figure \ref{fig:FP-dists}).

A comparison of the different histograms in each panel of Fig. \ref{fig:galFP-dists} shows that the distribution of FPs is mass-dependent whith more massive galaxies having relatively larger fractions of LOSs with shorter FPs. Consequently, the mean FP of all LOS at a given distance becomes smaller for more massive halos. This trend is stronger at smaller impact parameters and becomes less evident at large distances where the MFP of different mass bins become nearly identical (e.g., compare the top-left and bottom-right panels of Fig. \ref{fig:galFP-dists}.)

The sensitivity of the MFP measurements to halo mass and to the starting distance from the galaxies is shown in Figure \ref{fig:GalMFP} for the \emph{L25N512} simulation at $z = 8.5$ (right) and $z = 6$ (left). The three different halo mass bins used in the left panel are identical to those used in Figure \ref{fig:galFP-dists} where the blue solid, green dashed and red dotted curves show the MFP around galaxies, $\rm{MFP_{gal}}$, with mean halo masses $\left<\rm{M_{200} / M_{\odot}}\right>  = 10^{11}$, $\left< {\rm{M_{200} / M_{\odot}}}\right>  = 10^{10}$ and $\left<{\rm{M_{200} / M_{\odot}}} \right> =10^{9}$, respectively. To show the typical scatter in the distribution of FPs, the 15th-85th percentiles are shown for the highest halo mass bin using blue dot-dot-dashed curves. 

As discussed above, at $z = 6$ the $\rm{MFP_{gal}}$ increases with starting distance from galaxies for all three halo mass bins and converges to the MFP in the IGM (measured using randomly selected LOS) at distances $r \gtrsim 1$ cMpc away from the galaxies. At shorter distances, $\rm{MFP_{gal}}$ becomes more sensitive to the halo mass and decreases with increasing halo mass. The sensitivity of $\rm{MFP_{gal}}$ to halo mass is strongest at $r \lesssim 30$ ckpc  away from the galaxies. At those distances, however, $\rm{MFP_{gal}}$ is dominated by a few long FPs while most of the LOS have very short FPs. Therefore, the curve showing the 85th percentile of the FP distribution falls below the $\rm{MFP_{gal}}$ at $r < 30$ ckpc. Those small distances are likely to be closely associated with the ISM gas whose physical properties are not fully resolved in our cosmological simulations. Therefore, one should be cautious when interpreting the simulation results at those relatively small scales.

At $z = 8.5$ (where the neutral fraction is $\approx 0.5$) $\rm{MFP_{gal}}$ grows steeply with increasing distance at $r \lesssim 30$ ckpc and even exceeds the IGM MFP for $r \gtrsim 100$ ckpc. This reflects the fact that before reionization, regions around galaxies are more ionized than an average location in the IGM. At large enough distances from the galaxies (at $\gtrsim 1$ cMpc), however, the starting point of the $\rm{MFP_{gal}}$ calculation drops outside of those ionized bubbles and the $\rm{MFP_{gal}}$ starts to decrease with increasing distance. The dependence of $\rm{MFP_{gal}}$ on halo mass reverses at $r \sim 20$ ckpc: while at smaller distances more massive halos have shorter $\rm{MFP_{gal}}$, in the distance range $20 \lesssim r \lesssim 2000$ ckpc the $\rm{MFP_{gal}}$ is longer around more massive halos. This trend shows that at $z = 8.5$ the ionized bubbles tend to be larger and more ionzed around more massive galaxies. The aforementioned trends are consistent with inside-out reionization at $z = 8.5$, when the ionized regions surrounding galaxies have not yet merged.

\section{Summary and conclusions}
\label{sec:dend}
In this paper we used the Aurora radiation-hydrodynamical simulations to study the mean free path for ionizing photons during the epoch of reionization. The large dynamic range of physical scales available in the Aurora simulation suite, the rich collection of physical processes included in them, together with their consistency with a wide range of observational constraints on the properties of galaxies and the IGM, make the Aurora simulations suitable for tackling this problem. 
We focused on the nature and evolution of the mean free path of 1 Ry hydrogen ionizing photons as a measure of the opacity of the IGM. By using randomly selected lines-of-sight through our simulations, we directly measured the distance 1 Ry photons travel before reaching an optical depth of unity along each LOS (FP). Combining large numbers of LOS, we investigated different statistical properties of those distances, including their mean, i.e., the mean free path (MFP), their physical properties and their correlation with galaxies.

We found that the MFP evolves by more than 3 orders of magnitude during reionization. It increases from MFP $\lesssim 10$ pkpc at $z \gtrsim 10$, when the IGM is almost completely neutral, to $\gtrsim 10$ pMpc at $z \lesssim 6$, when reionization is complete and the IGM $\HI$ fraction is $\lesssim 10^{-4}$. We also found that small differences between the MFP and its evolution in simulations with different box-sizes and resolutions are mainly due to small differences in the IGM neutral fraction in those simulations. Indeed, the evolution of the MFP as a function of IGM neutral fraction converges much better with the box size and resolution of the simulations. This also applies to simulations with very different reionization histories. We showed that in a simulation with late reionization, where the volume-weighted $\HI$ fraction of $0.5$ is reached at $z \approx 7$ instead of $z \approx 8.3$, the MFP can be lower than in our reference model by more than 2 orders of magnitude at fixed redshifts. However, the MFPs in the two simulations are nearly identical at fixed IGM $\HI$ fractions (see Fig. \ref{fig:MFP-evol}).

We showed that all our simulations agree with the observational measurements of the MFP available for $z \approx 5~-~6$. However, the best agreement is produced by our reference model with a Thompson optical depth of $\tau_{\rm{reion}} \approx 0.068$ towards the CMB, which is consistent with the \citet{Planck15}. On the other hand, our late reionization model with $\tau_{\rm{reion}} = 0.055$, which is closer to the most recent Planck measurement \citep{Planck16}, slightly underproduces the observed MFP at $z \approx 5$ (see Fig. \ref{fig:MFP-evol}, left panel).

We found a relatively flat distribution of FPs along different LOS during the early stages of reionization. This results in a large scatter in the measured FPs. At later times, when the IGM $\HI$ fraction drops below a few percent, the distribution of FPs becomes less flat, with a peak slightly above the MFP, and an extended tail towards smaller values (see Fig. \ref{fig:FP-dists}). 

We also investigated the distribution of $\HI$ fractions and gas densities at the ODU points, which are defined as the locations along the different LOSs where the optical depth reaches unity (see Fig. \ref{fig:HI-nH-dists}). The distribution of the $\HI$ fraction at the ODU points peaks at nearly unity at early times and during reionization, but shifts to values $\lesssim 10^{-3}$ after the completion of reionization. The distribution of gas densities at the ODU points, on the other hand, peaks at values just below the mean baryonic density of the universe during the early stages of reionization, but shifts to significantly higher densities after the completion of reionization. Those trends imply an increasing importance of discrete, over-dense but mostly ionized systems in driving the mean opacity of the IGM after reionization.

By identifying patches of gas with $\HI$ fractions above a threshold value, we investigated the physical properties of $\HI$ systems during and after reionization (see Fig. \ref{fig:LHI-evol}). We found that for a threshold of $\nHI / \nH > 0.01$, the typical proper size of the $\HI$ systems changes by about 2 orders of magnitude during reionization. The $\HI$ sizes start from large values comparable to our simulation sizes before reionization and drop to $\lesssim 10$ pkpc after the IGM $\HI$ fraction decreases below a few percent. After the completion of reionization, the $\HI$ sizes remain nearly constant. The typical $\HI$ column density of the same systems evolves from $\NHI \gtrsim 10^{20}\cmsq$ before reionization to $\NHI \sim 10^{18-19}\cmsq$ afterwards, with a rather large scatter at all times. We also found that the size of $\HI$ systems is insensitive to the resolution of our simulations, and always remains well above the typical sub-kpc resolutions we use. We show that this conclusion is valid even if we use the $\nHI / \nH > 0.5$ threshold to identify neutral systems with typical $\sim 1$ pkpc sizes and $\HI$ column densities $\gtrsim 10^{20}\cmsq$ (see Fig. \ref{fig:LHI-evol-def}). 

Furthermore, we found that after reionization (i.e., an IGM $\HI$ fraction below a few percent) the typical size of the $\HI$ systems is close to their local Jeans lengths. This extends the simple Jeans scaling argument suggested by \citet{Schaye01} for the relationship between the $\NHI$ and local density to redshifts as high as $z \approx 7$. At earlier times, however, there is large scatter in the relation between gas density and its $\HI$ fraction. In other words, during the early stages of reionization, the size of the $\HI$ systems is generally not set by their local physical properties and may not correlate well with the local Jeans length.

From the \HI column density distribution function we computed the contributions of different column density ranges to the differential opacity, $\kappa_{912} = \frac{d\tau_{912}}{dr}$. We found that $d\kappa_{912}/ d \log{\NHI}$ is maximum for $\HI$ absorbers with $\NHI \sim 10^{18}~\cmsq$ both during and after reionization. However, the cumulative contribution of systems with $\NHI \lesssim 10^{16.5}~\cmsq$ is sufficiently high to drive the MFP at $z \approx 6$ (see Fig. \ref{fig:Kappa}). The column density threshold for the systems whose cumulative opacity corresponds to the directly measured MFP increases with the mean IGM $\HI$ fraction and hence with redshift. For instance, at $z = 8.3$, when the IGM $\HI$ fraction is $\approx 0.5$, the cumulative opacity of systems with $\NHI \lesssim 10^{18.5}\cmsq$ corresponds to the directly measured IGM MFP at that redshift (see Fig. \ref{fig:KappaEvol}).

To investigate the enhancement in the absorption of hydrogen ionizing photons close to galaxies, we calculated the mean free path starting from locations close to galaxies, $\rm{MFP_{gal}}$, instead of starting the calculation from randomly chosen points, as we did when calculating the IGM MFP. We found that close to galaxies $\rm{MFP_{gal}}$ becomes much smaller than the IGM MFP at the same redshift. This opacity enhancement is largest within $\approx 10-30$ ckpc from galaxies, and only becomes small at large distances, e.g., $r \gg 10^{-1}$ comoving Mpc for ${\rm{M_{200} \sim 10^{11} M_{\odot}}}$ at $z \approx 6$. This general picture holds at all redshifts. Before the completion of reionization, however, there is a reduction in the gas opacity at intermediate distances from galaxies. For instance, we found that at $z = 8.5$, $\rm{MFP_{gal}}$ at distances $D \sim 10^2-10^3$ ckpc is a few times longer than the IGM MFP. This is consistent with ISM gas (including that from sattelites) enhancing the opacity very near galaxies (i.e., $D \lesssim 10$ ckpc), while at larger distances and before entering the neutral gas far from galaxies, the opacity is reduced (i.e., $\rm{MFP_{gal}}$ is increased) within the ionized bubbles that preferentially form around galaxies during reionization. When reionization is complete and the ionized bubbles around galaxies have merged (e.g., $z = 6$), there is no reduction in the opacity at intermediate distances (see Fig. \ref{fig:GalMFP}). 

The enhanced opacity which results in a reduced MFP close to galaxies has important consequences for models that interpret the IGM MFP as the distance escaped ionizing photons can travel from galaxies before being absorbed by the IGM. Neglecting this opacity enhancement results in an over-estimate of the typical distances escaped ionizing photons can travel, and an under-estimate of the required escape fraction from galaxies, and/or the required emissivity of ionizing photons in the post-reionization Universe.

\section*{Acknowledgments}
\addcontentsline{toc}{section}{Acknowledgments}
We are grateful to Andreas Pawlik for leading the Aurora project and very useful discussions. We thank Nick Gnedin for fruitful discussions. Computer resources for this project have been provided by the Gauss Centre for Supercomputing/Leibniz Supercomputing Centre under grant: pr83le, the Partnership for Advanced Computing in Europe (PRACE) under proposal number 2013091919 and the Swiss National Supercomputing Centre (CSCS) under project ID s613. This work was sponsored with financial support from the Netherlands Organization for Scientific Research (NWO), also through VICI grant 639.043.409. We also benefited from funding from the European Research Council under the European Union's Seventh Framework Programme (FP7/2007- 2013)/ERC Grant agreement 278594-GasAroundGalaxies.

\end{document}

%% file: simparam.tex
\begin{table*}
\caption{List of Aurora simulations used in this work. The sub-resolution escape fraction for the first 5 simulations with different resolutions is calibrated such that they achieve the same reionization time (i.e., volume-weighted neutral fraction of 0.5 at $z \approx 8.3$) as explained in \citet{Pawlik16}. The \emph{L012N0256-L} simulation, on the other hand, uses a lower sub-resolution escape fraction resulting in later reionization (volume-weighted neutral fraction of 0.5 at $z \approx 7$). The supernovae feedback efficiency is different for different resolutions such that they all agree with the observed star formation rate function at $z \approx 7$.}
\begin{tabular}{lrrcccll}
\hline
Simulation & $L$       & $N$ & $m_{\rm b}$ & $m_{\rm dm}$ & $\epsilon_{\rm com}$ &  Remarks\\  
                & (cMpc/h) &       & $(\Msun)$     & $(\Msun)$        & (ckpc/h)                &  \\
\hline 
\emph{L25N512} &    25 & $2\times512^3$ & $2.04 \times 10^6$ & $ 1.00 \times 10^7$ & 1.95 & reference\\
\emph{L12N256} &    12 & $2\times256^3$ & $2.04 \times 10^6$ & $ 1.00 \times 10^7$ & 1.95 & small box\\
\emph{L12N512} &    12 & $2\times512^3$ & $2.55 \times 10^5$ & $ 1.25 \times 10^6$ & 0.98 & high resolution\\
\emph{L25N256} &    25 & $2\times256^3$ & $1.63 \times 10^7$ & $ 8.20 \times 10^7$ & 3.91 & low resolution\\
\emph{L50N512} &    50 & $2\times512^3$ & $1.63 \times 10^7$ & $ 8.20 \times 10^7$ & 3.91 & large box\\
\emph{L12N256-L} &    12 & $2\times256^3$ & $2.04 \times 10^6$ & $ 1.00 \times 10^7$ & 1.95 & late reionization\\
\hline
\end{tabular}
\label{tbl:sims}
\end{table*}